%% file: main.tex
\documentclass[acmtog]{acmart} 

\usepackage{booktabs}

\input{macros}

\setcopyright{rightsretained}
\acmJournal{TOG}
\acmYear{2022}\acmVolume{41}\acmNumber{4}\acmArticle{138}\acmMonth{7} \acmDOI{10.1145/3528223.3530157}



\begin{document}

\title{GANimator: Neural Motion Synthesis from a Single Sequence}

\author{Peizhuo Li}
\affiliation{\institution{ETH Zurich}  \country{Switzerland}}
\email{peizli@ethz.ch}

\author{Kfir Aberman}
\affiliation{\institution{Google Research} \country{USA}}
\email{kfiraberman@gmail.com}

\author{Zihan Zhang}
\affiliation{\institution{The University of Chicago} \country{USA}}
\email{zzhang18@uchicago.edu}

\author{Rana Hanocka}
\affiliation{\institution{The University of Chicago} \country{USA}}
\email{ranahanocka@gmail.com}

\author{Olga Sorkine-Hornung}
\affiliation{\institution{ETH Zurich} \country{Switzerland}}
\email{sorkine@inf.ethz.ch}

\begin{abstract}
\input{abstract}
\end{abstract}

\input{figures/000_teaser.tex}

\maketitle

\input{intro}
\input{relatedworks}
\input{method}
\input{experiments}
\input{application}
\input{conclusion}
\input{acks}

\bibliographystyle{ACM-Reference-Format}
\bibliography{bibs}

\input{appendix}

\end{document}

%% file: macros.tex
\usepackage{hyperref}
\usepackage{amsmath}
\usepackage{graphicx}
\usepackage{comment}
\usepackage{multirow,bigdelim}
\usepackage{algpseudocode}
\usepackage{algorithm}

\usepackage{array}
\usepackage{wrapfig}

\usepackage{csvsimple}
\usepackage{adjustbox}
\usepackage[percent]{overpic}

\DeclareMathOperator*{\argmin}{arg\,min}
\newcommand{\etal}{et al.}

\newcommand{\bbo}{{\bf O}}
\newcommand{\bbq}{{\bf Q}}

\newcommand{\bbr}{{\bf R}}
\newcommand{\bbs}{{\bf S}}
\newcommand{\bbt}{{\bf T}}
\newcommand{\bbl}{{\bf L}}
\newcommand{\bbc}{{\bf C}}
\newcommand{\bigG}{\mathcal{G}}
\newcommand{\bbe}{\mathbb{E}}
\newcommand{\bbp}{\mathbb{P}}
\newcommand{\Jfoot}{\mathcal{F}}
\newcommand{\Loss}{\mathcal{L}}
\newcommand{\Motion}{\mathcal{M}}
\newcommand{\JConstrain}{\mathcal{C}}
\newcommand{\Gen}{G}
\newcommand{\Dis}{D}
\newcommand{\FK}{\text{FK}}

\usepackage{dsfont}
\newcommand{\R}{\mathds{R}} 

\usepackage{color}
\usepackage{soul}
\usepackage{breqn}
\usepackage{booktabs}
\usepackage{tabularx}
\usepackage{rotating}

\definecolor{purple}{rgb}{0.99,0.2,0.72}
\definecolor{blue}{rgb}{0, 0.2, 0.8}
\definecolor{orange}{rgb}{0.6, 0.6, 0}
\definecolor{red}{rgb}{0.8, 0.2, 0.2}
\definecolor{magenta}{rgb}{0.5, 0.0, 1.0}
\definecolor{black}{rgb}{0.0, 0.0, 0.0}
\definecolor{cyan}{rgb}{0, 0.65, 0.65}
\definecolor{olive}{rgb}{0.2, 0.6, 0.5}
\definecolor{jackGreen}{RGB}{0,79,12}
\usepackage{xcolor}

\newif\ifdraft
\drafttrue

\ifdraft
\newcommand{\rhc}[1]{{\color{magenta}[\textbf{Rana:} \textit{#1}]}}
\newcommand{\kac}[1]{{\color{orange}[\textbf{Kfir:} \textit{#1}]}}
\newcommand{\plc}[1]{{\color{blue}[\textbf{Peizhuo:} \textit{#1}]}}
\newcommand{\OSHc}[1]{{\color{purple}[\textbf{Olga:} \textit{#1}]}}
\newcommand{\jzh}[1]{{\color{jackGreen}[\textbf{Jack:} \textit{#1}]}}

\else
\newcommand{\rhc}[1]{}
\newcommand{\kac}[1]{}
\newcommand{\plc}[1]{}
\newcommand{\OSHc}[1]{}
\newcommand{\jzh}[1]{}

\fi

\usepackage{pgfplots}
\pgfplotsset{compat=newest}
\usepgfplotslibrary{groupplots}
\usepgfplotslibrary{dateplot}

\usepackage[bottom]{footmisc}
\newcommand{\secref}[1]{Sec.~\ref{#1}}

\newcommand{\figref}[1]{Fig.~\ref{#1}}

\newcommand{\eqnref}[1]{Eq.~(\ref{#1})}


%% file: abstract.tex
We present GANimator, a generative model that learns to synthesize novel motions from a single, short motion sequence. GANimator generates motions that resemble the core \textit{elements} of the original motion, while simultaneously synthesizing novel and diverse movements. Existing data-driven techniques for motion synthesis require a large motion dataset which contains the desired and specific skeletal structure.
By contrast, GANimator only requires training on a \emph{single} motion sequence, enabling novel motion synthesis for a variety of skeletal structures \textit{e.g.,} bipeds, quadropeds, hexapeds, and more. Our framework contains a series of generative and adversarial neural networks, each responsible for generating motions in a specific frame rate. The framework progressively learns to synthesize motion from random noise, enabling hierarchical control over the generated motion content across varying levels of detail. We show a number of applications, including crowd simulation, key-frame editing, style transfer, and interactive control, which all learn from a single input sequence. Code and data for this paper are at \url{https://peizhuoli.github.io/ganimator}.

%% file: figures/000_teaser.tex
\begin{teaserfigure}
    \newcommand{\traincolor}{\color[rgb]{0.26,0.64,0.17}}
    \newcommand{\testcolor}{\color[rgb]{0.15,0.53,0.87}}
    \centering
    \includegraphics[width=\linewidth]{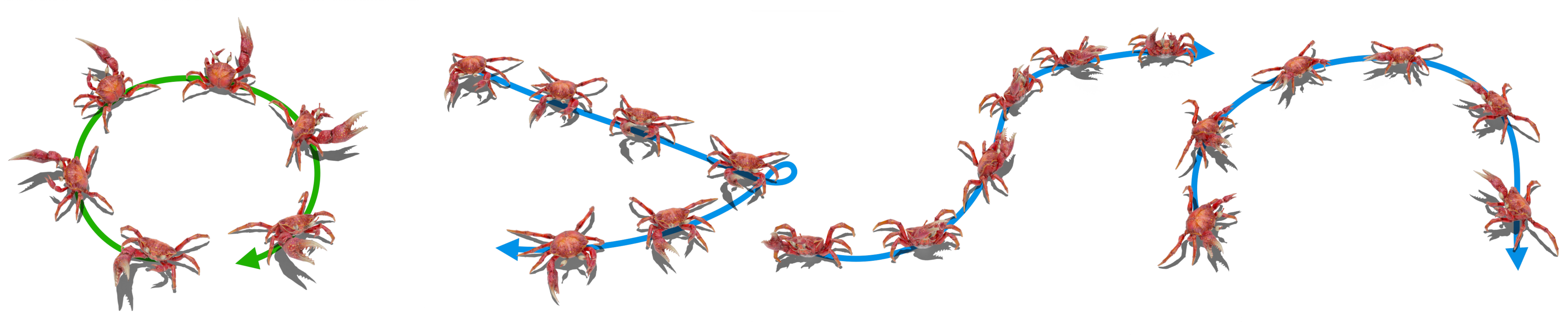}
    \caption{Given {\traincolor \textbf{a single training motion sequence}} of skeleton with arbitrary topology, our method learns to {\testcolor \textbf{synthesize novel motion sequences}}.}
    \label{fig:teaser}
\end{teaserfigure}

%% file: intro.tex
\section{Introduction}
Generating realistic and diverse human motion is a long-standing objective in computer graphics.
Motion modeling and synthesis commonly uses a probabilistic model to capture limited local variations~\cite{li2002motion}  
or utilizes a large motion dataset obtained by motion capture (mocap) {\cite{holden2016deep}}. 
Capturing data with a mocap system is costly both during stage-setting and in post-process (e.g., involving manual data clean-up).
Motion datasets are often limited, i.e., they lack the desired skeletal structure, body proportions, or styles. Therefore, utilizing motion datasets often requires non-trivial processing such as retargeting, which may potentially degrade or introduce errors in the original captured motion.
Moreover, there are no extensive datasets which contain imaginary creatures or non-standard animals (such as the hexapedal crab in Figure~\ref{fig:teaser}), which limits existing data-driven techniques. 

In this work, we develop a framework that is capable of generating diverse and realistic motions using only a \emph{single} training sequence. Our strategy greatly simplifies the data collection process while still allowing the framework to create realistic variations and faithfully capture the details of the individual motion sequence.
Inspired by the SinGAN model for image synthesis~\cite{shaham2019singan}, our key idea is to leverage the information within a single motion sequence in order to exploit the rich data over multiple temporal and spatial scales.
The generation process is divided into several levels, similar to progressive training schemes for images~\cite{karras2018progressive, shaham2019singan}.

Our framework is an effective tool for generating novel motion which is not exactly present in the single given training sequence. GANimator can synthesize long, diverse, and high-quality motion sequences using only a short motion sequence as input. We show that for various types of motions and characters, we generate sequences that still resemble the core \textit{elements} of the original motion, while simultaneously synthesizing novel and diverse movements. It should be noted that the synthesized motion is {not} copied verbatim from the original sequence. Namely, patches from the original motion do not appear in the synthesized motion.

We demonstrate the utility of the GANimator framework to generate both \emph{conditional} and \emph{unconditional} motion sequences that are \emph{reminiscent} of the input motion. GANimator can unconditionally (\textit{i.e.,} driven by noise) generate motions for {simulating crowds} and {motion editing/mixing}. In addition, GANimator is able to produce \textit{controllable} motion sequences, which are conditioned on a user-given input (\textit{e.g.,} trajectory position). We are able to inject global position and rotation
of the root joint as an input for interactively controlling the trajectory of the synthesized motion. 
GANimator is also able to perform {key-frame editing}, which generates high-quality interpolations between user-modified poses. Lastly, GANimator enables {motion style transfer} -- synthesizing the style of the input motion onto a different motion.

Our system relies on skeleton-aware operators~\cite{aberman2020skeleton} as a backbone for our neural motion synthesis framework. The skeleton-aware layers provides the platform for applying convolutions over a fixed skeleton topology. Since our network trains on a single motion sequence, we automatically adjust the operators to adhere to the structure of the input skeleton. Therefore, our system is able to train on a wide variety of skeletal structures, \textit{e.g.,} bipeds, quadropeds, hexapeds, and more. Further, for ground-inhabiting creatures, we incorporate a self-supervised foot contact label. This ensures proper placement of the feet on the ground plane and avoids notorious foot sliding artifacts. We demonstrate the effectiveness of GANimator in handling a wide variety of skeletal structures and motions, and its applicability in various motion editing and synthesis tasks.
 
Achieving desirable results in this constrained scenario is highly challenging, with existing techniques commonly producing undesirable results that fall into one of two extremes. In the first extreme, generated results span the breadth of poses contained in the original motion sequence, but are jittery and incoherent. In the second extreme, results are smooth, but lack variety and coverage of the motion elements contained in the original sequence. Our proposed technique strikes a favorable balance between these extremes, synthesizing high-quality, novel, and varied motion sequences. Our framework produces desirable motion sequences that contain all the original motion elements, while still achieving diverse and smooth motion sequences.

%% file: relatedworks.tex
\section{Related Work}

We review various relevant works on motion generation, focusing on human and other character animation. For an in-depth survey of this extensive body of literature we refer the readers to the surveys in \cite{PhysicsCharacterAnimationSTAR:2011,mourot2021survey}.

\paragraph{Statistical modeling of visual and motion data.}
Representing and generating complex visual objects and phenomena of stochastic nature, such as stone and wood textures or breathing and walking cycles with natural variations, is one of the fundamental tasks in computer graphics.
In one of the pioneer works in this area, Perlin~\shortcite{perlin1985image} introduced the famous Perlin noise function that can be used for image synthesis with variations. This  technique was later applied to procedural animation synthesis~\cite{perlin1996improv} to achieve random variability. Instead of using additive noise, statistical approaches~\cite{pullen2000animating, brand2000style, bowden2000learning, chai2007constraint} propose using probabilistic models such as kernel-based distribution and hidden Markov models, enabling synthesis with variations via direct sampling of the model.
{To capture both local and global structure, MotionTexture~\cite{li2002motion} clusters a large training dataset into textons, a linear dynamics model capable of capturing local variation, and models the transition probability between textons. Though the goal resembles ours, this data-hungry statistical model does not fit in our single-motion setting. We perform an in-depth comparison to MotionTexture regarding the ability to model both global and local variations when data is limited in \secref{sec:main_compare}. A similar idea of using two-level statistical models is employed in multiple works~\cite{tanco2000realistic, lee2002interactive} for novel motion synthesis.}
Other statistical models based on Gaussian processes learn a compact latent space~\cite{grochow2004style, wang2007gaussian, levine2012continuous} or facilitate an interactive authoring process~\cite{ikemoto2009generalizing} from a small set of examples for multiple applications, including inverse kinematics, motion editing and task-guided character control. \citet{lau2009modeling} use a similar Bayesian approach to model the natural variation in human motion from several examples. However, probabilistic approaches tend to capture limited variations in local frames and require a large dataset for a better understanding of global structure variations. 

\paragraph{Interpolation and blending of examples.}
Another approach for example-based motion generation is explicit blending and concatenation of existing examples. Early works~\cite{rose1996efficient, wiley1997interpolation, rose1998verbs, mizuguchi2001data, park2002line} use various interpolation and warping techniques for this purpose. Matching-based works~\cite{pullen2002motion, arikan2002interactive, kovar2004automated} generate motion by matching user-specified constraints in the existing dataset and interpolating the retrieved clips. Kovar \etal~\shortcite{kovar2002motion} explicitly model the structure of a corpus of motion capture data using an automatically constructed directed graph (the \emph{motion graph}). This approach enables flexible and controllable motion generation and is widely used in interactive applications like computer games. Min \etal~\shortcite{min2012motion} propose a mixture of statistical and graph-based models to decouple variations in the global composition of motions or actions, 
and the local movement variations. One significant constraint of blending-based methods is that the diversity of the generated results is limited to the dataset, since it is composed of interpolations and combinations of the dataset. Further works~\cite{heck2007parametric, safonova2007construction, zhao2009automatic} focus on the efficiency and robustness of the algorithm but are still inherently limited in the variety of the generated results. Motion matching~\cite{buttner2015motion} directly finds the best match in the mocap dataset given the user-provided constraints, bypassing the construction of a graph to achieve better realism. However, it generally requires task-specific tuning and is unable to generate novel motion that are unseen in the dataset.

\paragraph{Physics-based motion synthesis.}
Another loosely related field is physically-based motion generation, where the generation process runs a \emph{controller} in a physics simulator, unlike kinematics approaches that directly deal with joints' transformations.
\citet{agrawal2013diverse} optimize a procedural controller given a motion example to achieve physical plausibility while ensuring diversity in generated results. However, such controllers are hand-crafted for specific tasks, such as jumping and walking, and are thus difficult to generalize to different motion data. Results in~\cite{ye2010synthesis, wei2011physically} show that combining physical constraints and statistical priors helps generate physically realistic animations and reactions to external forces, but the richness of motion is still restricted to the learned prior. 
With the evolution of deep reinforcement learning, control policies on different bodies including biped locomotion~\cite{heess2017emergence,peng2017deeploco} and quadurpeds~\cite{luo2020carl} can be learned from scratch without reference. It is also possible to achieve high quality motion by learning from reference animation~\cite{peng2018deepmimic}.
\citet{lee2021learning} propose to learn a parameterized family of control policies from a single clip of a single movement, e.g., jumping, kicking, backflipping, and are able to generate novel motions for a  different environment, target task, and character parameterization. As in other works, the learned policy is limited to several predefined tasks. 

\paragraph{Neural motion generation.}
\citet{taylor2009factored} made initial attempts to model motion style with neural networks by restricted Boltzmann machines. Holden \etal~\shortcite{holden2015learning,holden2016deep} applied convolutional neural networks (CNN) to motion data for learning a motion manifold and motion editing. Concurrently, \citet{fragkiadaki2015recurrent} chose to use recurrent neural networks (RNN) for motion modeling. RNN based works also succeed in short-term motion prediction~\cite{fragkiadaki2015recurrent, pavllo2018quaternet}, interactive motion generation~\cite{lee2018interactive} and music-driven motion synthesis~\cite{aristidou2021rhythm}. \citet{zhou2018auto} tackle the problem of error accumulation in long-term random generation by alternating the network's output and ground truth as the input of RNN during training.
This method, called acRNN, is able to generate long and stable motion similar to the training set. However, like many other deep learning based methods, it struggles when only a short training sequence is provided, whereas we are able to address the limited data problem. We compare our method to acRNN in \secref{sec:main_compare}.
\citet{holden2017phase} propose phase-functioned neural networks (PFNN) for locomotion generation and introduce \emph{phase} to neural networks. Similar ideas are used in quadruped motion generation by \citet{zhang2018mode}. \citet{starke2020local} extend phase to local joints to cope with more complex motion generation. \citet{henter2020moglow} propose another generative model for motion based on normalizing flow. Neural networks succeed in a variety of motion generation tasks: motion retargeting~\cite{villegas2018neural, aberman2020skeleton, aberman2019learning}, motion style transfer~\cite{aberman2020unpaired, mason2022real}, key-frame based motion generation~\cite{harvey2020robust}, motion matching~\cite{holden2020learned} and animation layering~\cite{starke2021neural}. It is worth noting that the success of deep learning methods hinges upon large and comprehensive mocap datasets. 
However, acquiring a dataset involves costly capturing steps and nontrivial post-processing, as discussed in Section 1. {In contrast, our method achieves comparable results by learning from a single input motion sequence.} 

%% file: method.tex
\section{Method}

We propose a generative model that can learn from a single motion sequence. Our approach is inspired by recent works in the image domain that use progressive generation  \cite{karras2018progressive} and works that  propose to train deep networks on a single  example~\cite{shocher2019ingan,shaham2019singan}. 
We next describe the main building blocks of our hierarchical framework, motion representation, and the training procedure.

\subsection{Motion representation}
We represent a motion sequence by a temporal set of $T$ poses that consists of root joint displacements $\bbo \in \R^{T \times 3}$ and joint rotations $\bbr \in \R^{T \times JQ}$, where $J$ is the number of joints and $Q$ is the number of rotation features. The rotations are defined in the coordinate frame of their parent in the kinematic chain, and represented by the 6D rotation features ($Q = 6$) proposed by Zhou \etal~\shortcite{zhou2019continuity}, which yields the best result among other representations for our task (see ablation study in Section~\ref{sec:ablation}). 

To mitigate common foot sliding artifacts, we incorporate foot \emph{contact labels} in our representation. In particular, we concatenate to the feature axis $C\cdot T$ binary values $\bbl\in\{0,1\}^{T\times C}$, which correspond to the contact labels of the foot joints, $\Jfoot$, of the specific creature. For example, for humanoids, we use $\Jfoot = \{$left heel, left toe, right heel, right toe$\}$. For each joint $j\in\Jfoot$ and frame $t\in \{1, \ldots, T\}$, the $tj$-th label is calculated via 
$$\bbl^{tj} = \mathds{1}[\|\FK_\bbs(\left[\bbr, \bbo\right])^{tj}\|_2 < \epsilon],$$ 
where $\|\FK_\bbs(\left[\bbr, \bbo\right])^{tj}\|_2$ denotes the magnitude of the velocity of joint $j$ in frame $t$ retrieved by a forward kinematics (FK) operator.  FK applies the rotation and root joint displacements on the skeleton $\bbs$, and  $\mathds{1}[V]$ is an indicator function that returns 1 if $V$ is true and $0$ otherwise. 

To simplify the notation, we denote the metric space of the concatenated features by
$\Motion_T \equiv \R^{T \times(JQ+C+3)}$. In addition, we denote the input motion features by $\bbt \equiv [\bbr,\bbo,\bbl]\in \Motion_{T}$, and its correponding downsampled versions by $\bbt_i\in\Motion_{T_i}$.

\subsection{Progressive motion synthesis architecture}

\input{figures/004_train_arch.tex}

Our motion generation framework is illustrated in Fig~\ref{fig:train_arch}. It consists of $S$ coarse-to-fine generative adversarial networks (GANs)~\cite{goodfellow2014generative}, each of which is responsible to generate motion sequences with a specific number of frames $\{T_i\}_{i=1}^S$. We denote the generators and discriminators by $\{\Gen_i\}_{i=1}^S$ and $\{\Dis_i\}_{i=1}^S$, respectively. 

The first level is purely generative, namely, $\Gen_1$ maps a random noise $z_1 \in \Motion_{T_1}$ into a coarse motion sequence
\begin{equation}
    \bbq_1 =\Gen_1(z_1),
\end{equation}
where $\bbq_1\in\Motion_{T_i}$.
Then, the generators in the finer levels $\Gen_i$ $(2 \leq i \leq S)$ progressively upsample $\bbq_1$ via
\begin{equation}
    \bbq_i =\Gen_i(\bbq_{i-1}, z_i),
\end{equation}
where in each level the sequence is upsampled by a fixed scaling factor $F>1$. The process is repeated until the finest output sequence $\bbq_S \in \Motion_{T_S}$ is generated by $G_S$.

Note that $z_i \in \Motion_{T_i}$ has an i.i.d normal distribution $\sim\mathcal{N}\left(0, \sigma_i\right)$ along the temporal axis while being shared along the channel axis, and we found that $\sigma_i$ is highly correlated with the magnitude of the high-frequency details generated by $G_i$, thus, we select 
\begin{equation}
    \label{eq:sigma}
    \sigma_i = \frac{1}{Z_i} \|\uparrow\bbt_{i-1} - \bbt_i\|_2^2,
\end{equation}
where $Z_i=T_i(QJ+C+3)$ is the number of entries in $\bbt_i$, and $\uparrow$ is a linear upsampler with a scaling factor $F > 1$.
In all of our experiments we select $F = 4/3$ and $S=7$.

\subsubsection{Network components}

\paragraph{Generator} Our generator $G_i$ contains a fully convolutional neural network $g_i(\cdot)$ that has a few skeleton-aware convolution layers~\cite{aberman2020skeleton} followed by non-linear layers (see Appendix~\ref{sec:netarch}). Since the main role of the network is to add missing high-frequency details, we use a residual structure \cite{he2016deep}, hence for $2\leq i \leq S$, we get
\begin{equation}
    \label{eq:generator}
    \Gen_i(\bbq_{i-1}, z_i) = g_i(\uparrow\bbq_{i-1} + z_i) + \uparrow\bbq_{i-1}.
\end{equation}

\paragraph{Discriminator.} While discriminators in classic GAN architectures output a single scalar indicating whether the input is classified as ``real'' or ``fake'',
such a structure in the case of a single sequence in the training data leads to mode collapse, since our generator trivially overfits the sequence. In order to prevent overfitting we limit the receptive field of the discriminator by employing a Patch-GAN~\cite{isola2017image, li2016precomputed} classifier, which calculates a confidence value to each input patch. The final output of our discriminator is the average value of all the per-patch confidence values, predicted by the network. 

\paragraph{Skeleton-aware operators.} We employ skeleton-aware convolutions~\cite{aberman2020skeleton} as a fundamental building block in our framework. 
Skeleton-aware operators require a fixed skeleton topology that is defined by a set of joints (vertices) and an adjacency list (edges).
Since our network operates on a single sequence we adapt the topology to adhere the input sequence. This enables operation on any skeleton topology, and does not require retargeting the input motion to a specific skeletal structure. 
To incorporate the foot contact labels into the skeleton-aware representation, we treat each label as a virtual joint connected to its corresponding joint rotations vertex. In addition, due to the high correlation between end-effectors and global positions, we add a connection between the contact labels vertices to the vertex of the global displacements $\bbo$, where we treat the latter as another virtual joint that is connected to the neighbors of the root joint in the kinematic chain.

\subsubsection{Loss functions}
\paragraph{Adversarial Loss} We train level $i$ with the WGAN-GP~\shortcite{gulrajani2017improved} loss:
\begin{eqnarray}
    \Loss_{\text{adv}} &= &\bbe_{\bbq_i \sim \bbp_{g_i}} \left[ \Dis_{i}(\bbq_i) \right] -  D_i(\bbt_i) \\ \nonumber
     & + & \lambda_{\text{gp}} \bbe_{\hat \bbq \sim \bbp_{\hat g_i}} \left[ \left( \| \nabla D_i(\hat \bbq_i) \|_2 - 1 \right) ^2  \right],
     \label{eq:gp}
\end{eqnarray}
where $\bbp_{g_i}$ is defined as the distributions of our generated samples in the $i$th level, $\nabla$ is the gradient operator and $\bbp_{\hat g_i}$ is the distribution of the linear interpolations $\hat \bbq_i = \lambda\bbq_i + (1 - \lambda)\bbt_i$ with $\lambda$ as a uniformly distributed variable in $[0, 1]$. The gradient penalty term in~\eqref{eq:gp} enforces  Lipschitz continuity so the Wasserstein distance between generated and training distribution can be well approximated~\cite{gulrajani2017improved}.

\paragraph{Reconstruction Loss} To ensure that the network generates variations of all the different temporal patches, and does not collapse to generation of a specific subset of movements, we require the network to reconstruct the input motion from a set of pre-defined noise signals $\{z_i^*\}_{i=1}^S$, namely, $\Gen_i(\uparrow \bbt_{i-1}, z_i^*)$ should approximate the single training example $\bbt_i$ at level $i$. To encourage the system to do so, we define a reconstruction loss
\begin{equation}
    \label{eq:rec_loss}
    \Loss_{\text{rec}} = \| \Gen_i(\uparrow\bbt_{i-1}, z_i^*) - \bbt_i\|_1.
\end{equation}
Note that the noise models the variation of generated results, but during reconstruction, we do not expect any variation. To this end, we fix $z_1^*$ as a pre-generated noise for the first level and set $z_i^* = 0$ for the other levels.
 
\input{figures/005_single_level.tex}

\paragraph{Contact Consistency Loss} As accurate foot contact is one of the major factors of motion quality, we predict the foot contact labels in our framework and use an IK post-process to ensure the contact. Since the joint contact labels $\bbl$ are integrated into our motion representation $\Motion$, the skeleton-aware networks can directly operate on $\Motion$ and learn to predict the contact label as part of the motion.

We noticed that the implicit learning of contact labels can cause artifacts in the transition between activated and non-activated contact labels. Thus, we propose a new loss that encourages a consistency between contact label and feet velocity. We require that in every frame either the contact label or the foot velocity will be minimized, via
\begin{equation}
    \label{eq:con_loss}
    \Loss_{\text{con}} = \frac{1}{T|\Jfoot|}\sum_{j \in \Jfoot} \sum_{t=1}^T \|\FK_\bbs(\bbr, \bbo)^{tj}\|_2^2 \cdot s(\bbl^{tj}),
\end{equation}
where $s(x) = 1 / [1 + \exp(5 - 10x)]$ is the transformed sigmoid function. We demonstrate the effectiveness of this loss in the ablation study (Section~\ref{sec:ablation}).

\subsubsection{Training}
Our full loss used for training summarizes as:
\begin{equation}
    \label{eq:full_loss}
    \Loss = \lambda_{\text{adv}}\Loss_{\text{adv}} + \lambda_{\text{rec}}\Loss_{\text{rec}} + \lambda_{\text{con}}\Loss_{\text{con}}.
\end{equation}
Although each level can be trained separately, the generated samples of generators in the low levels may be over-blurred due to low temporal resolution and the smoothing effect applied by convolutional kernel. To improve the robustness and quality of the results, we combine every 2 consecutive levels as a block and train the framework block by block. A similar technique is also used by Hinz \etal~\shortcite{hinz2021improved}.

For a detailed description of the layers in each component and the specific values of the hyper-parameters, we refer the reader to Appendix~\ref{sec:netarch}.

%% file: figures/004_train_arch.tex
\begin{figure*}
    \centering
    \includegraphics[width=\linewidth]{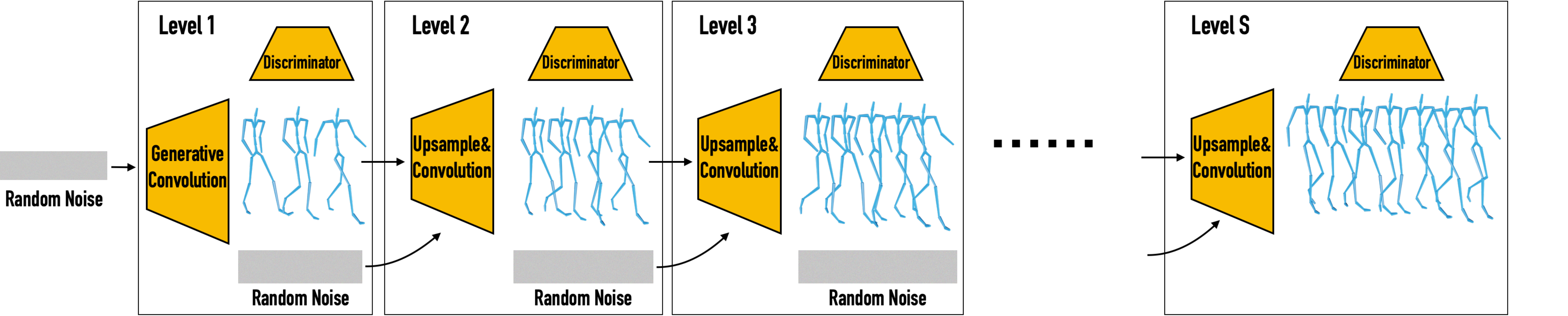}
    \caption{Our progressive motion synthesis architecture. Starting with random noise that synthesize a coarse motion sequence through a generative network, our framework progressively upsamples the motion until it reaches to the finest temporal resolution. Each level receives the output of the previous level (except for the first level) and a random noise as output an upsampled version of the input sequence. We use adversarial training, such that each generated result is fed into a discriminator at a corresponding level.}
    \label{fig:train_arch}
\end{figure*}

%% file: figures/005_single_level.tex
\begin{figure}
    \centering
    \includegraphics[width=\columnwidth]{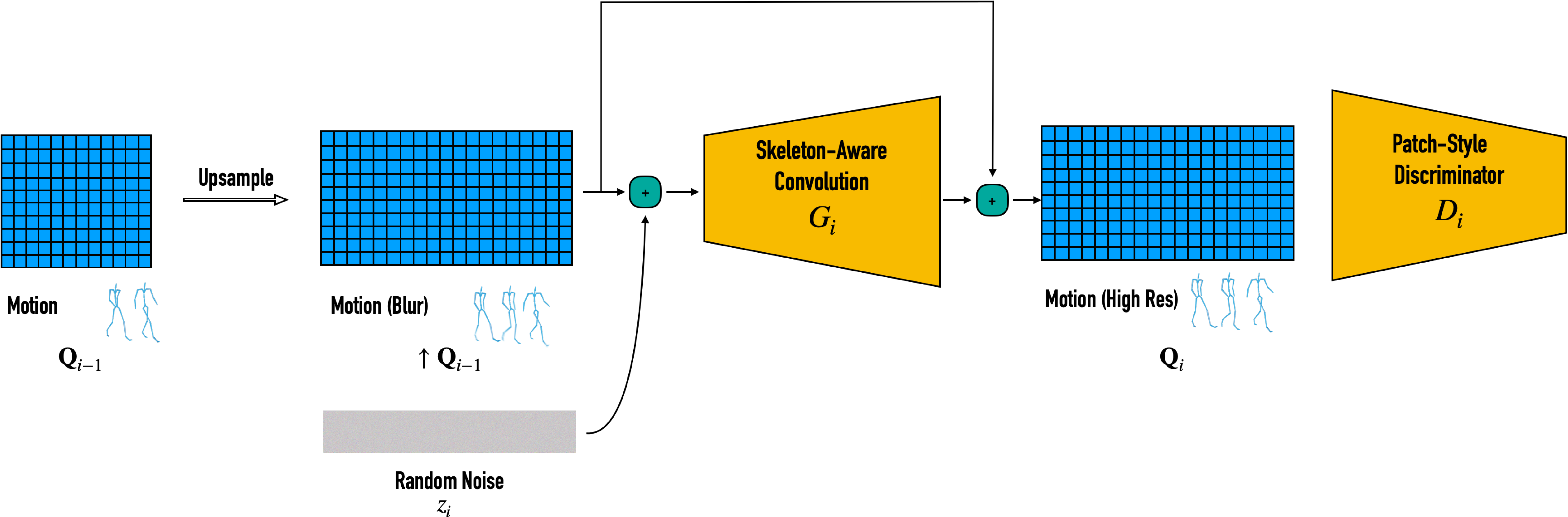}
    \caption{Generator structure. Our residual generator receives the sum of a noise vector and the upsampled result from previous level (in the first level it receives only noise). It predicts the missing high-frequency details, which is added to the input animation via skip connections. }
    \label{fig:single_level}
\end{figure}

%% file: experiments.tex
\section{Experiments and Evaluation}

We evaluate our results, compare them to other motion generation techniques and demonstrate the effectiveness of various components in our framework through an ablation study. Please refer to the supplementary video for the qualitative evaluation.

\subsection{Implementation details}

Our GANimator framework is implemented in PyTorch~\cite{NEURIPS2019_9015}, and the experiments are performed on NVIDIA GeForce RTX 2080 Ti GPU. We optimize the parameters of our network with the loss term in \eqnref{eq:full_loss} using the Adam optimizer~\cite{kingma2014adam}. We used different training sequences, whose length ranges from 140 frames to 800 frames at 30fps. It contains both artist-created animation and motion capture data from Mixamo~\shortcite{mixamo} and Truebones~\shortcite{truebones}. The training time is proportional to the training sequence length, e.g., it takes about 4 hours to train our network on a human animation sequence with around 600 frames (15,000 iterations per level).

\subsection{Novel motion synthesis}
\label{sec:main_compare}

\input{figures/010_vanilla.tex}

We demonstrate the ability of motion sequence extrapolation in \figref{fig:vanilla} and compare our method to the recent acRNN work \cite{zhou2018auto} and the classical statistical model MotionTexture~\cite{li2002motion}.
We quantitatively compare these methods with metrics dedicated to a single training example in Section~\ref{sec:evaluation}. 

Since our network is fully convolutional, we can generate high-quality motion sequences of arbitrary length given a single training example. As a simple application, we can easily generate a crowd animation, as shown in \figref{fig:crowd_simulation} and the accompanying video.

When only a single training sequence is provided, acRNN can only generate a limited number of frames before converging to a constant pose, because the lack of data leads to an overfitted model that is not robust to perturbation and error accumulation in RNNs, while our fully convolutional framework does not suffer from this issue.

MotionTexture~\cite{li2002motion} automatically clusters the patches of the training dataset into several textons and models the transition probability between textons, where each texton represents the variation of a small segment of similar motions. MotionTexture relies on similar but not identical patches in the dataset for modeling local variations. It constructs the global transition probability between textons based on the frequencies of consecutive relationships of corresponding patches in the dataset.
However, when only a single training sequence is provided, we need to manually divide the sequence into patches and specify the transition between textons. When splitting the training sequences into several textons and manually permuting them to create global structure variations, the transitions between textons can be unnatural (see \figref{fig:vanilla}). It is possible to learn a single texton for the whole training sequence to achieve better quality, but the model creates almost zero local variations due to limited data.

\input{figures/012_crowd_simulation.tex}

\subsection{Evaluation}
\label{sec:evaluation}

We discuss quantitative metrics for novel motion synthesis from a single training sequence and compare our method and existing motion generation techniques.

\paragraph{Coverage.}

An important quantitative measurement for the quality of our model is the \emph{coverage} of the training example. Since there is only one training example, we measure the coverage on all possible temporal windows $\mathcal{W}(\bbt, L) = \left\{\bbt^{i\,:\,i+L-1}\right\}_{i=1}^{L_T - L + 1}$ of a given length $L$, where $\bbt^{i\,:\,j}$ denotes the sequence of frames $i$ to $j$ of the training example $\bbt$, and $L_T$ is the total length of $\bbt$. Given a generated result $\bbq$, we label a temporal window $\bbt_w \in \mathcal{W}(\bbt, L_c)$ as covered if its distance measure to the nearest neighbor in $\bbq$ is smaller than an empirically chosen threshold $\epsilon$. The coverage of animation $\bbq$ on $\bbt$ is defined as
\begin{equation}
    \label{eq:cov_seq2seq}
    \text{Cov}(\bbq, \bbt) = \frac{1}{|\mathcal{W}(\bbt, L_c)|} \sum_{\bbt_w \in \mathcal{W}(\bbt, L_c)} \mathds{1}\left[ \text{NN}(\bbt_w, \bbq) < \epsilon \right].
\end{equation}
$\text{NN}(\bbq_1, \bbq_2)$ denotes the distance of the nearest neighbor of the animation sequence $\bbq_1$ in $\bbq_2$. It is crucial to use an appropriate distance measure here. In our setting, we choose the Frobenius norm on the local joint rotation matrices:
\begin{equation}
    \label{eq:single_nn}
    \text{NN}(\bbq_1, \bbq_2) = \frac{1}{L_1} \min_{\bbq_w \in \mathcal{W}(\bbq_2, L_1)} \left\|\bbq_1 - \bbq_w\right\|_F^2,
\end{equation}
where $L_1$ is the length of $\bbq_1$. We use joint rotations and not positions since our model creates local variations so that location deviations accumulate along the kinematics chain, which would cause location-based high distance measures on visually similar patches.

We choose $L_c = 30$, capturing local patch length of 1 second.
Similarly, the coverage of a model $\bigG(\cdot)$ on $\bbt$ is defined by 
\begin{equation}
    \label{eq:cov_mod2seq}
    \text{Cov}(\bigG, \bbt) = \mathbb{E}_{z} \text{Cov}(\bigG(z), \bbt).
\end{equation}

\paragraph{Global diversity.} 

\input{figures/009_dp_nearest_neighbor.tex}

To quantitatively measure the global structure diversity against a single training example, we propose to measure the distance between patched nearest neighbors (PNN). The idea is to divide the generated animation into several segments, where each segment is no shorter than a  threshold $T_\min$, and find a segmentation that minimizes the average per-frame nearest neighbor cost as illustrated in Figure~\ref{fig:dp_nn}.
For each frame $i$ in the generated animation $\bbq$ we match it with frame $l_i$ in the training animation $\bbt$, such that between every neighboring points in the set of discontinuous points $\{i\ |\ l_i \neq l_{i-1} +1\}$ is at least $T_\min.$ This is because every discontinuous point corresponds to the starting point of a new segment. We call such an assignment $\{l_i\}_{i=1}^L$ a \emph{segmentation} on $\bbq.$ When large global structure variation is present, it is difficult to find a close nearest neighbor for large $T_\min$. The patched nearest neighbor is defined by, minimizing over all possible segmentations,
\begin{equation}
    \Loss_{\text{PNN}} = \min_{\{l_i\}} \frac{1}{L}\sum_{i=1}^L \|\bbq^i - \bbt^{l_i}\|_F^2,
    \label{eq:pnn}
\end{equation}
where $\bbq^i$ denotes frame $i$ in $Q$ and $\bbt^{l_i}$ denotes frame $l_i$ in $\bbt$. The PNN can be solved efficiently with dynamic programming,  similar to MotionTexture~\shortcite{li2002motion}; we refer the reader to Appendix~\ref{sec:dp_detail} for more details.
In all our experiments, we choose $T_{\min} = 30$, corresponding to one second of the animation.

\paragraph{Local diversity.}

Our framework synthesizes animations carrying similar but \emph{diverse} visual content compared to the training sequence. We measure the local frame diversity by comparing every local window $\bbq_w \in \mathcal{W}(\bbq, L_d)$ of length $L_d$ to its nearest neighbor in the training sequence $\bbt$:

\begin{equation}
    \label{eq:local_diverse}
    \Loss_{\text{local}} = \frac{1}{|\mathcal{W}(\bbq, L_d)|} \sum_{\bbq_w \in \mathcal{W}(\bbq, L_d)} \text{NN}(\bbq_w, \bbt).
\end{equation}
Similar to the definition of coverage, we use the Frobenius norm over local joint rotation matrices. We choose $L_d = 15$ to capture \emph{local} differences between the generated result and the training example.

We quantitatively compare our results to MotionTexture~\cite{li2002motion} and acRNN~\cite{zhou2018auto} using the metrics above as reported in Table~\ref{tab:main_cmp}. Since we use the nearest neighbor cost against the training sequence to measure diversity, high diversity score does not necessarily imply plausible results: MotionTexture creates unnatural transitions that are not part of the training sequence, and acRNN converges to a pose that does not exist in the training sequence also leading to a high diversity score.
It can be seen that acRNN has limited coverage due to its convergence to a static pose, while our method generates motions that cover the training sequence well. MotionTexture trained as a single texton overfits to the training sequence, creating little variation on both local and global scale. Meanwhile, our model strikes a good balance between generating plausible motions and maintaining diversity.

\begin{table}
    \caption{Quantitative comparison to existing motion generation techniques.}
    \begin{tabular}{l c c c }
    \toprule
    \small & \small Coverage $\uparrow$ & \small Global Diversity & \small Local Diversity\\
    \midrule
    \small MotionTexture~\shortcite{li2002motion}  &  84.6\%  & 1.03  & 1.04\\
    \small MotionTexture (Single) & 100\% & 0.21 & 0.33  \\
    \small acRNN~\shortcite{zhou2018auto} &  11.6\% & 5.63 & 6.69 \\
    \small Ours & 97.2\%  &  1.29 & 1.19  \\
    \bottomrule
    \end{tabular}
    \label{tab:main_cmp}
\end{table}

\subsection{Ablation study}
\label{sec:ablation}

We evaluate the impact of motion representation, temporal receptive field, and the foot contact consistency loss on our performance. The results are reported in Table~\ref{tab:ablation} and demonstrated in the supplementary video.

\paragraph{Reconstruction loss.} 
In this experiment, we discard the reconstruction loss and retrain our model. The supplementary video shows that the quality of the motion is degraded as a result. The reconstruction loss ensures that an anchor point in the latent space can be used to reconstruct the training sequence perfectly. Since our framework generates novel variations of the training sequence, this anchor point helps to stabilize the generated results. For the same reason, the reconstruction loss reflects the quality of generated results of our framework and we report the impact of different components on the quality of results in Table~\ref{tab:ablation}.

\begin{table}
    \caption{Quantitative comparison for ablation study.}
    \begin{tabular}{l c}
    \toprule
    \small & \small Rec. Loss $\downarrow$ \\
    \midrule
    \small W/o contact consist. loss & 4.04 \\
    \small Euler angle & 23.6 \\
    \small Quaternion  & 9.25 \\
    \small Full appraoch  &  2.85 \\
    \bottomrule
    \end{tabular}
    \label{tab:ablation}
\end{table}

\paragraph{Contact consistency loss.} In this experiment we discard the contact consistency loss $\Loss_{\text{con}}$ and retrain our model. Our framework predicts the foot contact label that can be used to fix sliding artifacts in a post-process. Although it can be learned implicitly, i.e., without the contact consistency loss, the generated result contains inconsistent global positions and contact labels, leading to unnatural leaning and transitions of contact status after the post-process. The accompanying video shows that the contact consistency loss promotes consistency between predicted animation and contact labels, providing a robust fix to sliding artifacts.

\paragraph{Rotation representation.} 
In this experiment we use three different representations of joint rotations to train our networks: Euler angles, quaternions, and 6D representation~\cite{zhou2019continuity}. The results are shown in the accompanying video. It can be seen that the network struggles to generate reasonable results with Euler angles because of the extreme non-linearity. Quaternions yield more stable results compared to Euler angles, but the double-cover problem and the non-linearity still cause sudden undesired changes and degrade the realism of the generated motion.

\paragraph{Temporal receptive field.} 
A key component of generating diverse and realistic motion with global variation is the choice of the temporal receptive field, which we explore in this experiment. We demonstrate the impact by training our framework on a clip of the YMCA dance in the accompanying video. We observe that when a large temporal receptive field is chosen, the network memorizes the global structure, creating a static pose in the middle of generation without being able to permute the dance sequence. When the temporal receptive field is small, the network observes only limited information about the context and generates jittering results. The network can generate smooth and plausible permutations of the dance when a suitable receptive field is picked.

%% file: figures/010_vanilla.tex
\begin{figure*}
    \centering
    \begin{tabular}{c}
        \includegraphics[width=0.35\linewidth]{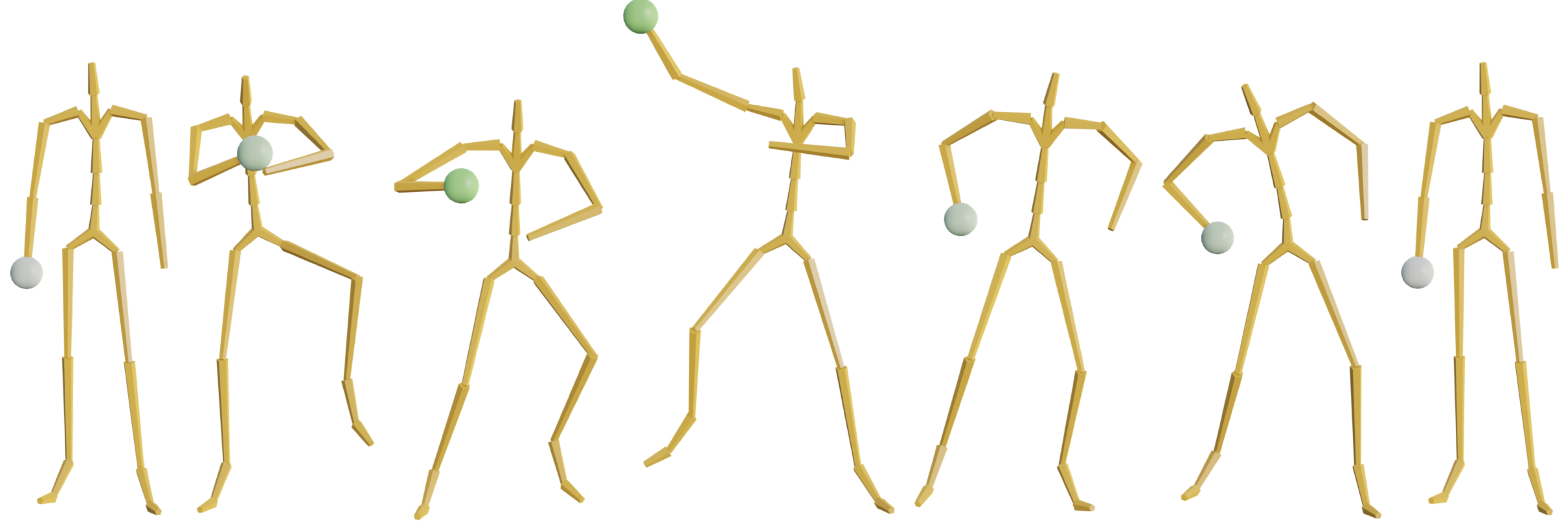} \\
        Training sequence\\
        \includegraphics[width=0.7\linewidth]{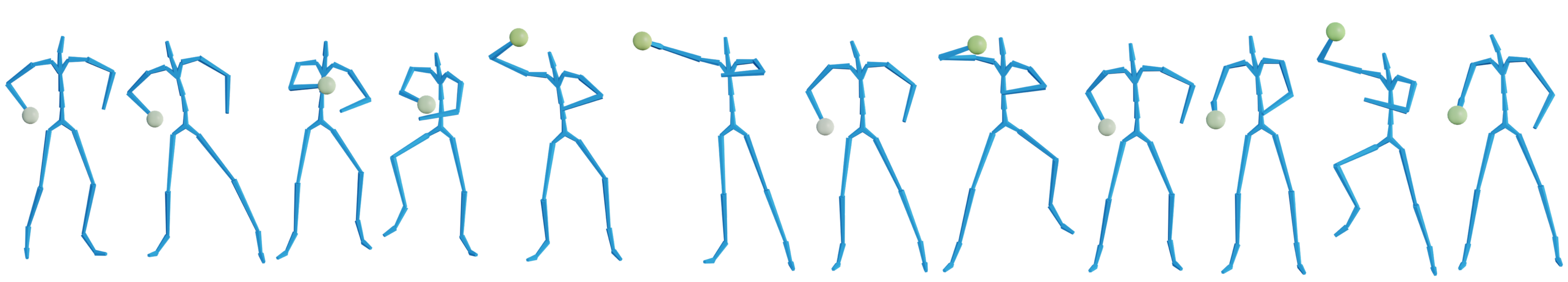} \\
        Ours\\
        \includegraphics[width=0.7\linewidth]{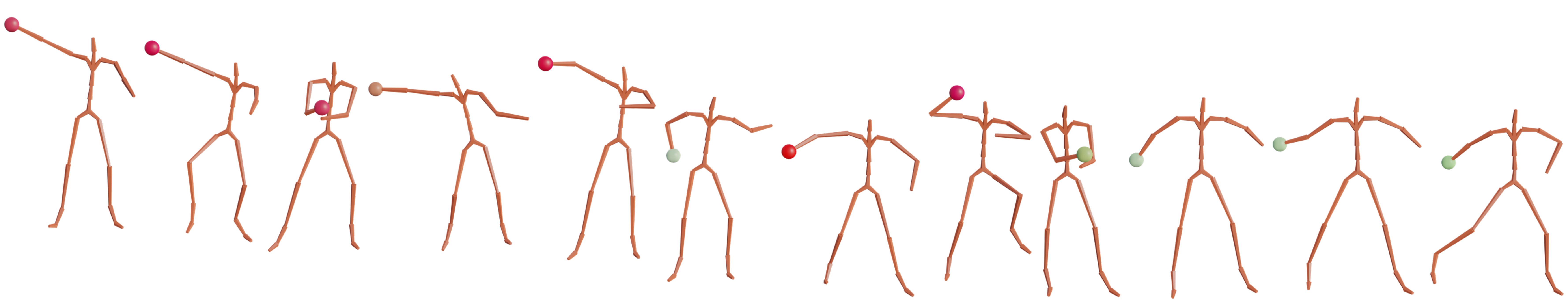} \\
        MotionTexture~\shortcite{li2002motion} \\
        \includegraphics[width=0.7\linewidth]{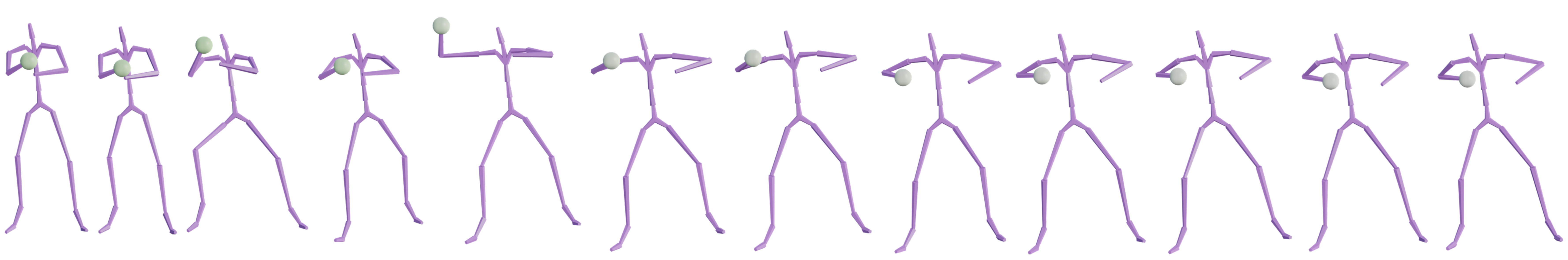} \\
        acRNN~\shortcite{zhou2018auto}\\
    \end{tabular}
    \caption{We train our framework, MotionTexture~\shortcite{li2002motion} and acRNN~\shortcite{zhou2018auto} on the Gangnam style dancing sequence with 371 frames and use them to synthesize a new sequence with 600 frames. The magnitude of velocity of the right hand is visualized with a heatmap (white - low, green - average, red - high). It can be seen that our method generates global structure variation, the poses and transitions look natural (see supplementary video) and visually similar to the training sequence. For MotionTexture~\shortcite{li2002motion}, we manually pick a path between all the trained textons to generate results with similar structure. However, it can be seen that such a process result in unnatural transitions visualized by the large hand velocity and bad foot contact. The result of acRNN~\shortcite{zhou2018auto} converges to a static pose very quickly due to insufficient data and thus the velocity of the hand gradually vanishes.}
    \label{fig:vanilla}
\end{figure*}

%% file: figures/012_crowd_simulation.tex
\begin{figure}
    \centering
    \begin{tabular}{c}
        \includegraphics[width=\columnwidth]{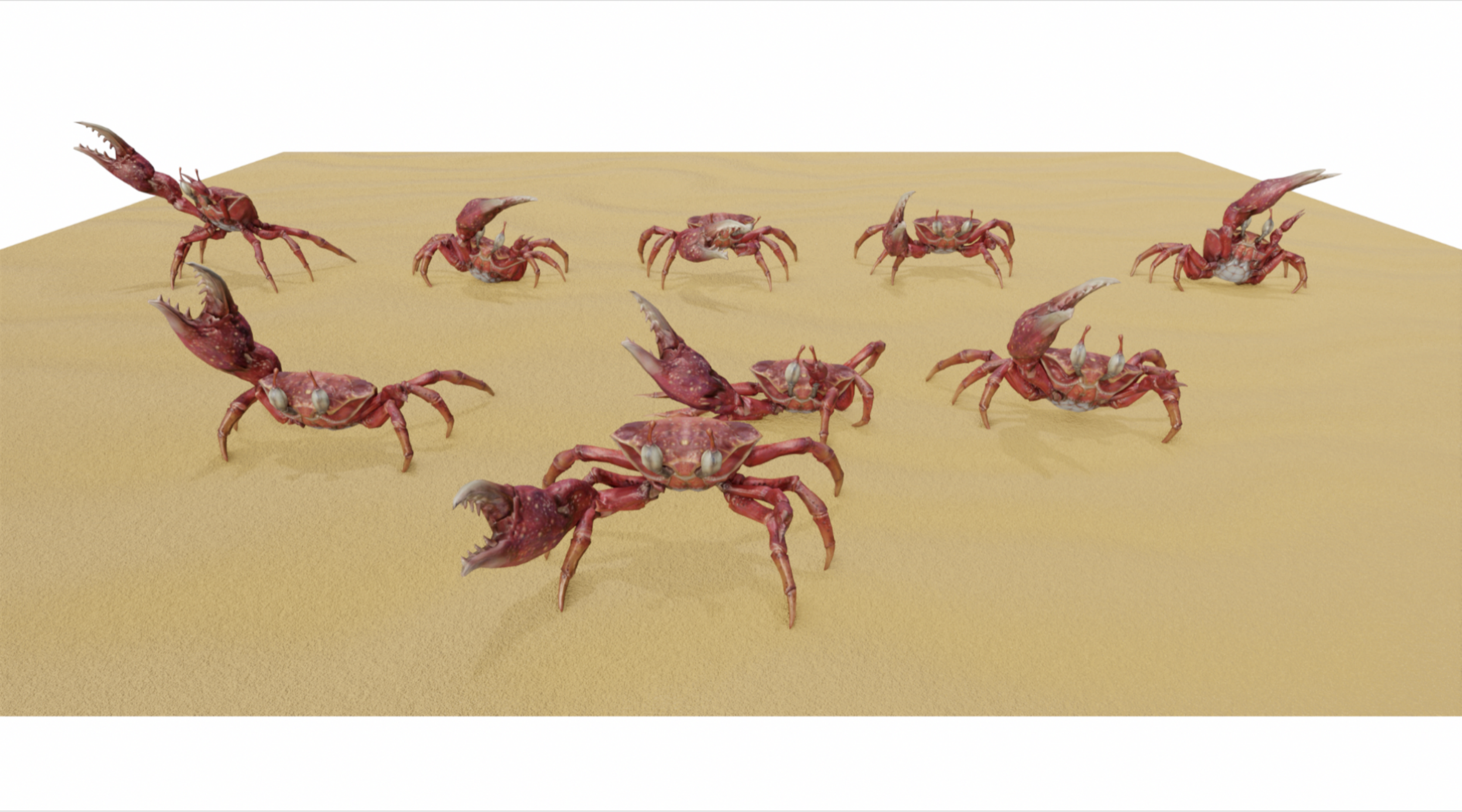} 
    \end{tabular}
    \caption{Crowd animation. Our framework trained on a single crab dancing sequence can synthesize various novel motions that can be used to simulate crowd and augment data for various purposes.}
    \label{fig:crowd_simulation}
\end{figure}

%% file: figures/009_dp_nearest_neighbor.tex
\begin{figure}
    \centering
    \includegraphics[width=\columnwidth]{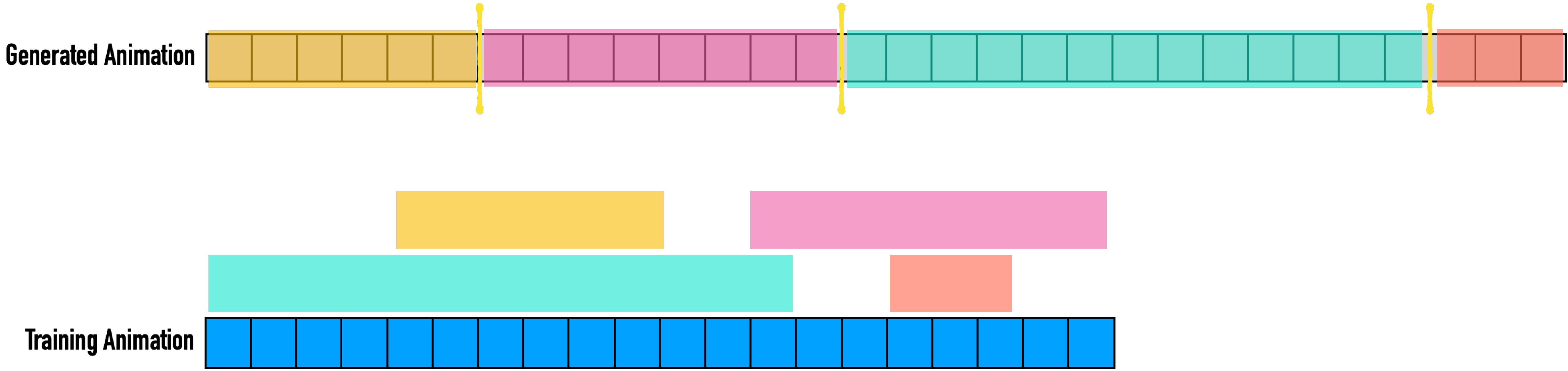}
    \caption{Patched Nearest Neighbor metric. The generated animation (top row) is divided into several segments, where the length of each segment is at least $T_{\min}$. Then each segment is assigned to its nearest neighbor in the training animation (bottom row) as visualised by the color bars.}
    \label{fig:dp_nn}
\end{figure}

%% file: application.tex
\section{Applications}

In this section we showcase the utility of pure generation and user-guided generation based on a single training example in different applications, including motion mixing, style transfer, key-frame editing, and interactive trajectory control.

\paragraph{Motion mixing} In addition to novel motion variation synthesis based on a single example animation, our framework can be also trained on several animations. Given $N$ training animations $\{\bbt^k\}_{k=1}^N$, we can train a single generative network on all $N$ examples by replacing the reconstruction loss in \eqnref{eq:rec_loss} with
\begin{equation}
    \label{eq:multi_rec}
    \Loss_{\text{rec}}' = \frac{1}{N} \sum_{k=1}^N \| \Gen_i(\uparrow\bbt^k_{i-1}, z_i^{k*}) - \bbt^k_i\|_1.
\end{equation}

We train our model on two animation sequences of an elephant, shown in \figref{fig:motion_mixing} and the video. We quantitatively measure the coverage of novel motion synthesis on two training animations in Table~\ref{tab:mixing_coverage}. It can be seen that the generated result of mixed training covers all the training examples well, and the synthesized output naturally fuses the two training sequences.
\begin{table}
    \caption{Coverage measurement of motion mixing.}
    \begin{tabular}{l r r}
    \toprule
    \small Training Set& \small Coverage of Seq. 1 & \small Coverage of Seq. 2\\
    \midrule
    \small Only Seq. 1 & 100.0\% & 1.5\% \\
    \small Only Seq. 2 & 2.4\% & 100\% \\
    \small All Seq. & 100.0\% & 97.8\%\\
    \bottomrule
    \end{tabular}
    \label{tab:mixing_coverage}
\end{table}

\input{figures/013_motion_mixing.tex}

\paragraph{Style transfer}
Our framework can perform motion style transfer using an input animation sequence whose style is applied to the content of another input animation. We exploit the hierarchical control of the generated content at different levels of detail. Since style is often expressed in relatively high frequencies, we use the downsampling of content input $\bbt^{C}$ to control the generation of the neural network trained on style input $\bbt^{S}$ for the style transfer task. Namely, we downsample the content animation $\bbt^{C}$ to the corresponding coarsest resolution $\bbt^{C}_1$ and use it to replace the output of the first level of the network trained on $\bbt^{S}$. We demonstrate our result in \figref{fig:style_transfer} and the video. It can be seen that, with a single example, we successfully transfer the proud style, yielding a proud walk sharing the same pace as the content input. Due to the fact that the output is generated by multi-scale patches learned from the style input $\bbt^{S}$, the content of $\bbt^{C}$ is required to be similar to the content of $\bbt^{S}$, in order to generate high-quality results (e.g., both of the inputs are walking).

\input{figures/014_style_transfer.tex}

\paragraph{Key-frame editing}
Our framework can also be applied to key-frame editing (\figref{fig:keyframe_editing}). We train our network on the input animation $\bbt$. The user can then manually edit some frames by changing their poses at the coarsest level $\bbt_1$, and the network produces smooth, realistic, and highly-detailed transitions between the modified key-frames, including unseen but plausible content.

\input{figures/015_keyframe_editing}

\paragraph{Conditional generation.}
Our framework can generate motion while accounting for user-specified constraints on the motion of selected joints.
Formally, given the constrained joints $\JConstrain$, the user-specified constraints $\bbc$ are given by $\bbq^\JConstrain = \{\bbq^j,\ {j \in \JConstrain}$, where $\bbq^j$ denotes the motion of joint $j$. In the case of motion generation with trajectory control, we set $\JConstrain = \{\text{root position, root orientation}\}$. 

\input{figures/006_conditional_single_level.tex}

The key component of enabling conditional generation is enforcing correlation between the generated motion and the constraints. The training process of the conditional generation is described in \figref{fig:cond_single_level}. As constraints are defined as part of the motion, we use the \emph{concatenation trick}, denoted by $concat(\bbq, \bbc)$, during the training and inference time, where $concat(\bbq, \bbc)$ is the result of replacing the motion of constrained joints $\bbq^\JConstrain$ in motion $\bbq$ by the constraints $\bbc$. 

For the evaluation of the $i$-th level, given the constraints at corresponding frame rate $\bbc_i$, we use $concat(\uparrow\bbq_{i-1}, \bbc_i)$ as the input of the generator instead of the upsampled result $\uparrow\bbq_{i-1}$ from the previous level. For training, we randomly generate $\bbc_i$ by sampling from a pre-trained generator mentioned above, taking the corresponding part of the motion and downsampling it to the corresponding frame rate. We employ the concatenation trick again on the output of the generator and let $\bbq_i = concat(\Gen_i(concat(\uparrow\bbq_{i-1}, \bbc_i), z_i), \bbc_i)$ as the output of this stage and feed it into the discriminator. The discriminator rules out motions that do not belong to the plausible motion distribution and forces the network to generate motions complying with the given constraints.

\paragraph{Interactive generation.}

\input{figures/007_interactive_generation.tex}

Conditional generation is particularly interesting for interactive applications, such as video games, where the motion of a character needs to be generated online in accordance to joystick controller inputs, for example.
We demonstrate the process for exploiting a pre-trained conditional generation framework for such interactive application in \figref{fig:interactive_gen}. Unlike RNNs that are straightforward to use for interactive or real-time generation, convolutional neural networks generally require modifications, such as causal convolution~\cite{oord2016wavenet}. We take a different approach by exploiting the limited receptive field of convolution.

Let $\bigG(\bbc, Z)$ be the multi-level conditional generation result with constraints $\bbc$, where $Z$ is the set of random noise $\{z_i\}_{i=1}^S$ used during generation. In the interactive setting, we assume there are existing constraints $\bbc_1$ and the corresponding random noise $Z_1$, which generates the result $\bbq_1 = \bigG(\bbc_1, Z_1)$, as demonstrated in \figref{fig:interactive_gen}(a). Given the new constraints $\bbc_2$ and corresponding noise $Z_2$, we concatenate them with $\bbc_1$ and $Z_1$ along the temporal axis, denoted by $\bbc = ext(\bbc_1, \bbc_2)$ and $Z = ext(Z_1, Z_2)$, and generate a new result $\bbq = \bigG(\bbc, Z)$. Denote the temporal receptive field of $\bigG(\cdot)$ by $R$ and the lengths of constraints $\bbc_1$ and $\bbc_2$ by $L_1$ and $L_2$, respectively. Note that ${\bbq}^{1\,:\,L_1-r}$ and $\bbq_1^{1\,:\,L_1-r}$ are equal (blue area in \figref{fig:interactive_gen}(b)), where $\bbq^{i\,:\,j}$ denotes frame $i$ to $j$ of motion $\bbq$ and $r = \lceil R/2 \rceil$ is the halved receptive field. The sequence ${\bbq}^{L_1 - r+ 1 \,:\, L_1}$ is different from $\bbq_1^{L_1 - r + 1 \,:\, L_1}$ and creates a smooth transition to the new constraints (dark cyan area in \figref{fig:interactive_gen}(b)). The remaining part ${\bbq}^{L_1 + 1 \,:\, L_1 + L_2}$ is the newly generated result complying with the new constraints $\bbc_2$ (cyan area in \figref{fig:interactive_gen}(b)). 

Therefore, given the new constraints, we only need to run the convolutional network on the concatenation of the last $2r$ frames of the existing constraints and the new constraints, as demonstrated by the activated area in \figref{fig:interactive_gen}(c), and withhold the last $r$ frames from displaying on the screen. The full method for interactive generation is summarized in Algorithm~\ref{alg:interactive_gen}. We demonstrate interactive trajectory control in \figref{fig:trajectory_control} and in the accompanying video.

\input{figures/016_trajectory_control}

\begin{algorithm}
    \caption{Interactive Generation}
    \label{alg:interactive_gen}
    \begin{algorithmic}
        \State $\bbc_1 \gets$ initial constraints
        \State $Z_1 \gets$ inital generated noise
        \State $r \gets$ halved receptive field of $\bigG(\cdot)$
        \State $\bbq \gets \bigG(\bbc_1, Z_1)$
        \State \emph{Display} $\bbq^{1\,:\,\text{end} - r}$
        \While{$\bbc_2 \gets$ new constraints}
            \State $Z_2 \gets$ generated noise
            \State $\bbc \gets ext(\bbc_1, \bbc_2)$
            \State $Z \gets ext(Z_1, Z_2)$
            \State $\bbq \gets \bigG(\bbc, Z)$
            \State \emph{Display} $\bbq^{r+1\,:\,\text{end} - r}$
            \State $\bbc_1 \gets \bbc^{\text{end}- 2r + 1 \,:\, \text{end}}$
            \State $Z_1 \gets Z^{\text{end} - 2r + 1 \,:\, \text{end}}$
        \EndWhile
    \end{algorithmic}
\end{algorithm}

%% file: figures/013_motion_mixing.tex
\begin{figure}
    \centering
    \begin{tabular}{c}
        \includegraphics[width=\columnwidth]{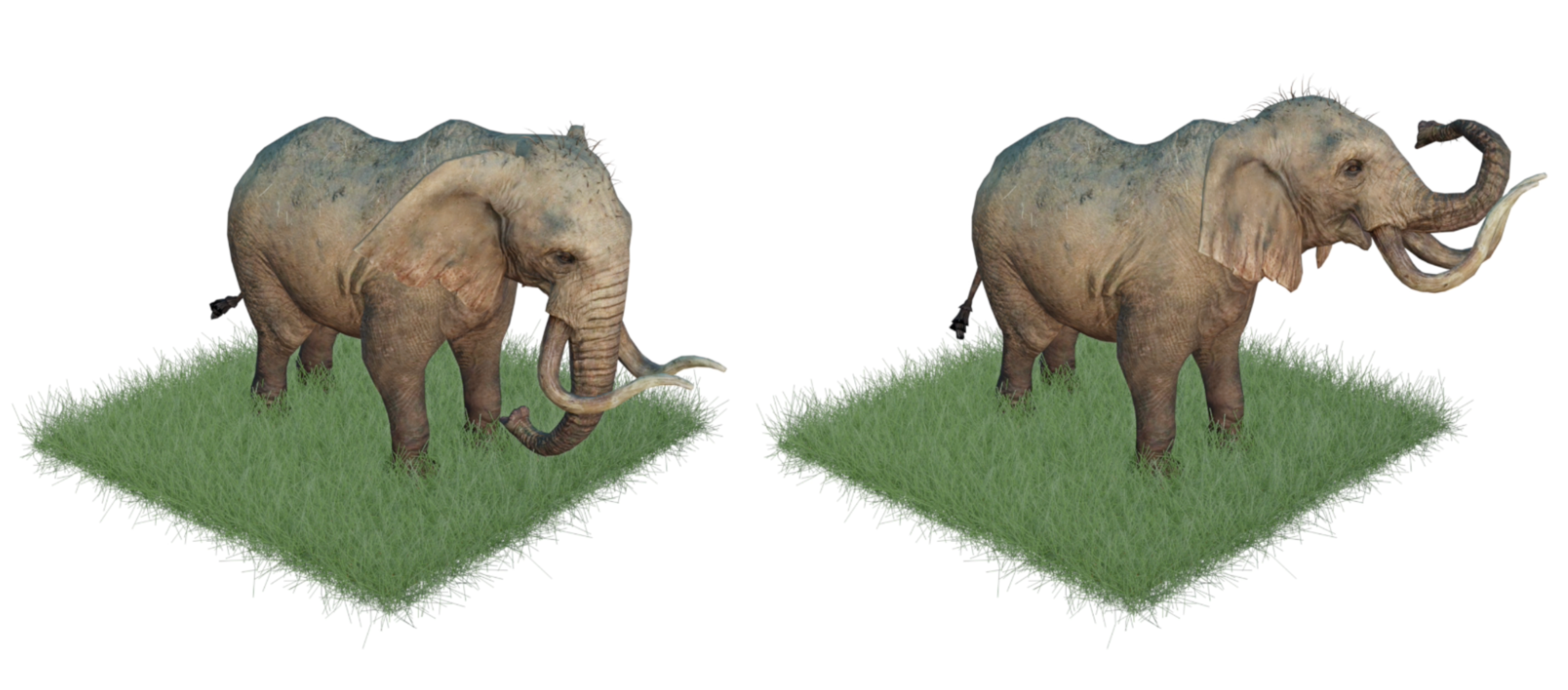}\\
        Training sequence 1\qquad \qquad Training sequence 2\\
        \includegraphics[width=\columnwidth]{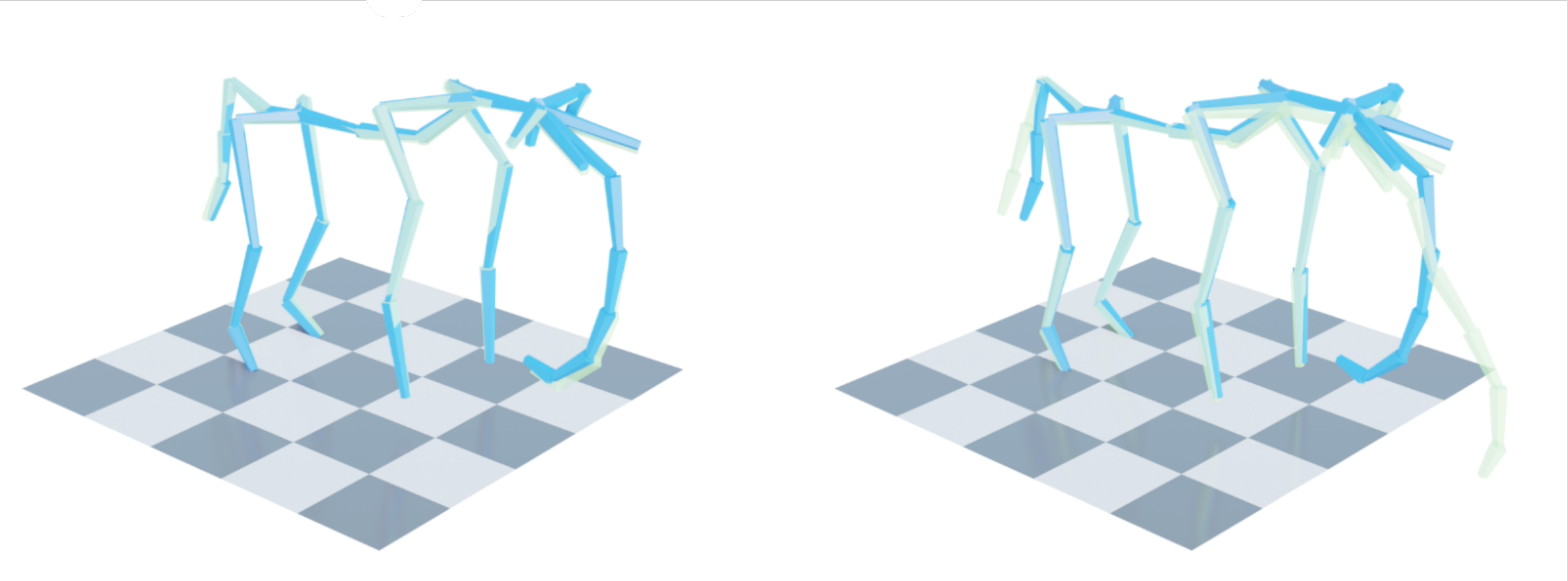}\\
        Closer to sequence 1\\
        \includegraphics[width=\columnwidth]{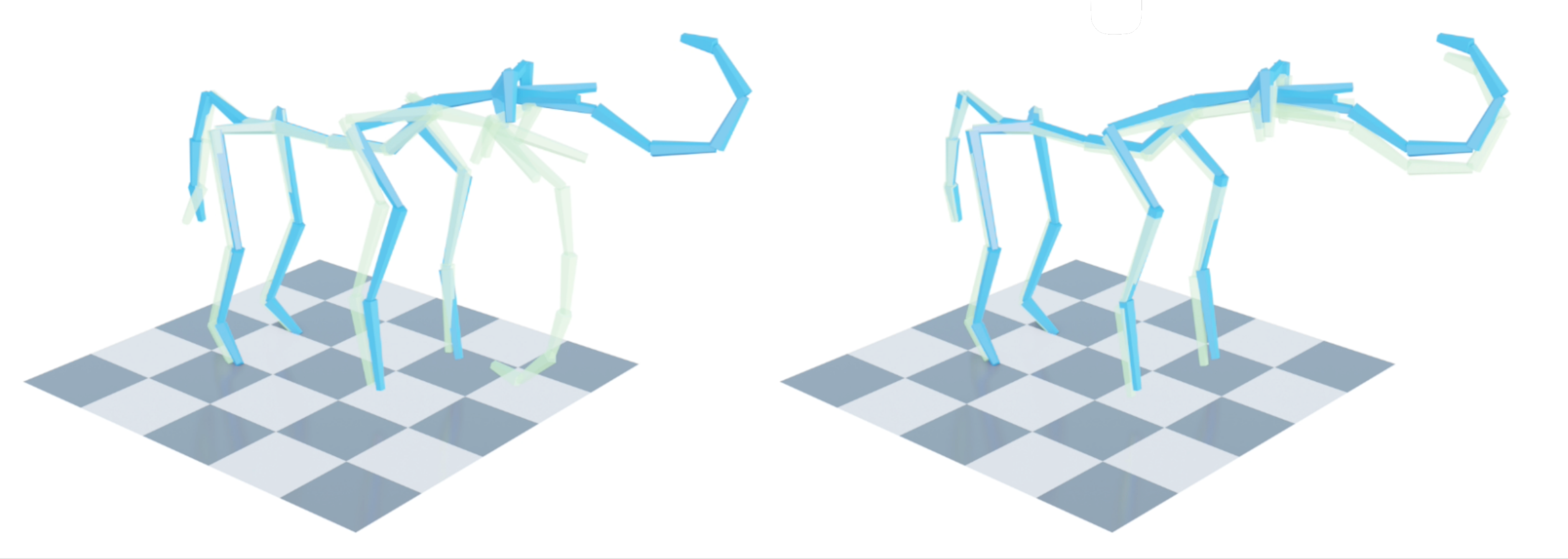}\\
        Closer to sequence 2\\
    \end{tabular}
    \caption{The model is trained with two sequences. The first sequence (left) contains relative static motion, the second sequence (right) contains larger movement. We visualize the skeletal animation of our generated result (blue) and its patched nearest neighbor (green) in corresponding sequences.  It can be seen that our result contains the content from both training sequence. }
    \label{fig:motion_mixing}
\end{figure}

%% file: figures/014_style_transfer.tex
\begin{figure}
    \centering
    \newcommand{\pll}{-4}
    \begin{overpic}[width=\columnwidth]{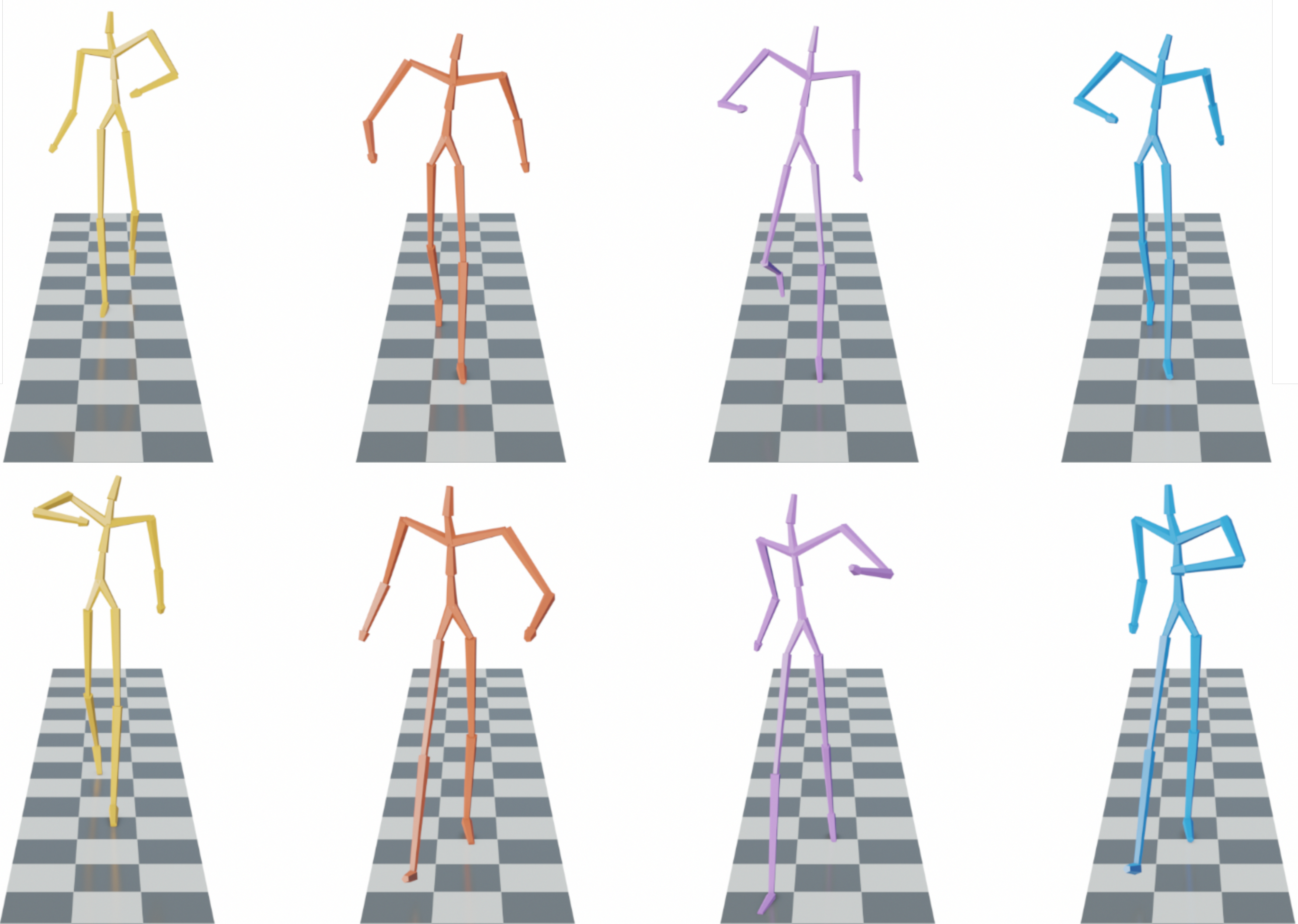}
    \put(-2,  \pll){\textcolor{black}{Style (proud)}}
    \put(30,  \pll){\textcolor{black}{Content}}
    \put(51,  \pll){\textcolor{black}{Aberman~\shortcite{aberman2020unpaired}}}
    \put(88, \pll){\textcolor{black}{Ours}}
    \end{overpic}
    \setlength{\abovecaptionskip}{10pt}
    \caption{We synthesize the proud style from the style input onto the content input. It can be seen that our result contains the same content as the content input while express the proud style, \textit{e.g.} higher elbow position on walking. We also show the result from ~\cite{aberman2020unpaired} for comparison.}
    \label{fig:style_transfer}
\end{figure}

%% file: figures/015_keyframe_editing.tex
\begin{figure}
    \centering
    \begin{tabular}{c}
        \includegraphics[width=\columnwidth]{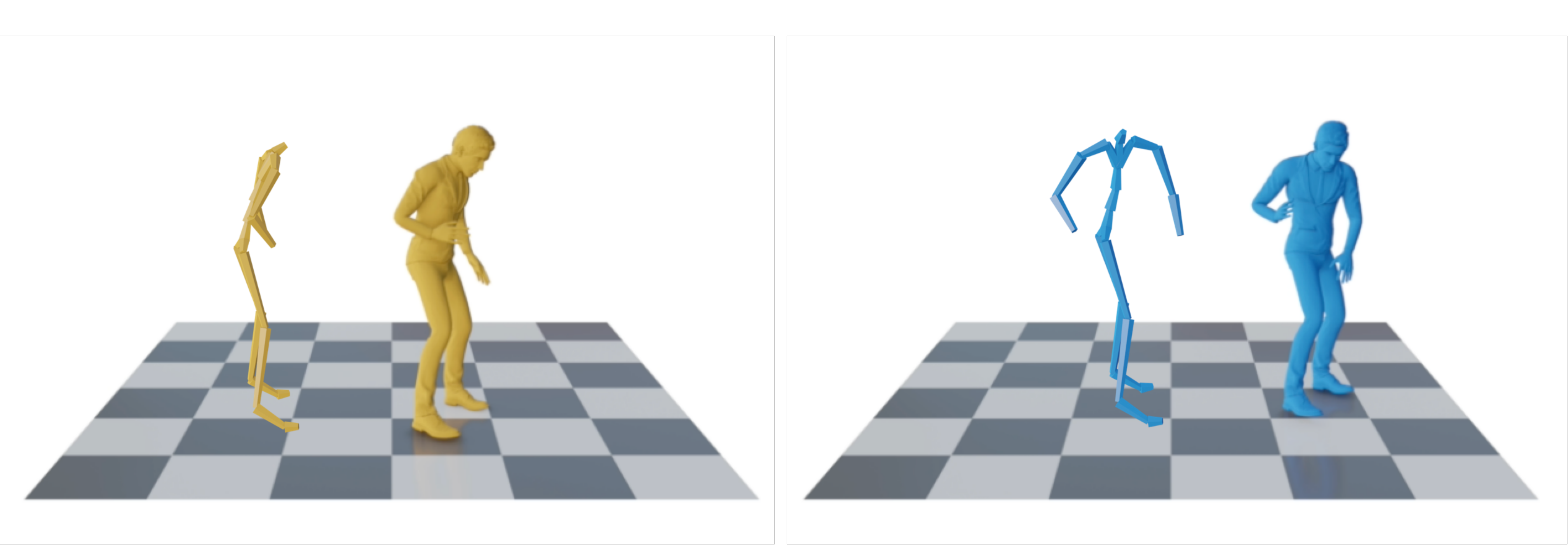}\\
        \put(-65,  0){\textcolor{black}{Input}}
        \put(50,  0){\textcolor{black}{Ours}}\\
        \includegraphics[width=\columnwidth]{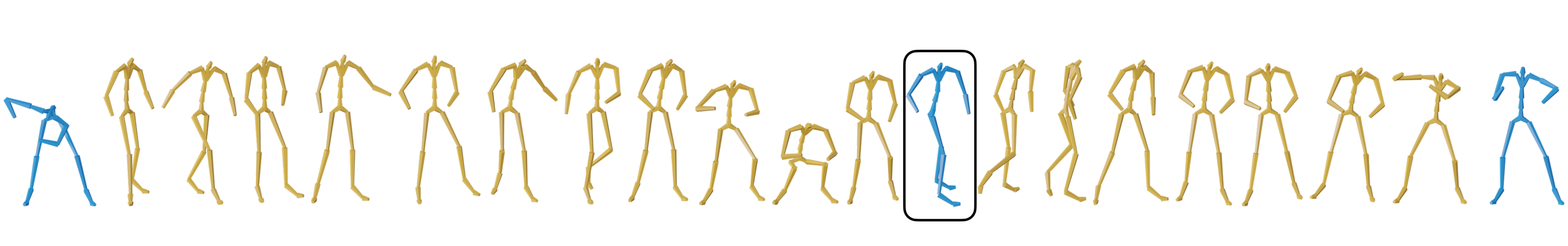}\\
        Edited Key-frames
        \\
    \end{tabular}
    \caption{We manually edit three key-frames (blue) in the input sequence by changing the poses. We visualize the result of one editing, where the character is made to face the audience. It can be seen that our model follows the editing and generates a plausible result. Please refer to the accompanying video for a complete result.}
    \label{fig:keyframe_editing}
\end{figure}

%% file: figures/006_conditional_single_level.tex
\begin{figure}
    \centering
    \includegraphics[width=\columnwidth]{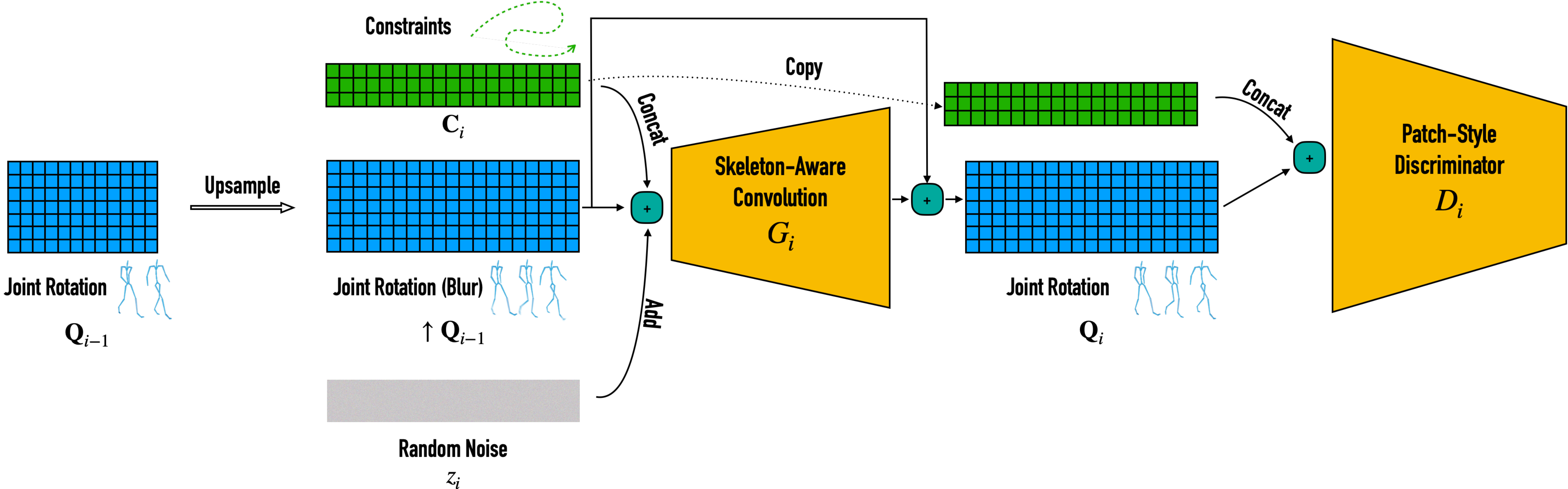}
    \caption{Training of conditional generation. To construct the input for the conditional generator, we concatenate the constraints (e.g., root joint movement) sampled from a pre-trained generator to the upsampled result generated with the same condition from the previous level. We again concatenate the constraints to the result produced by the generator and feed it to the discriminator. The discriminator ensures the correlation between constraints and the generated result.}
    \label{fig:cond_single_level}
\end{figure}

%% file: figures/007_interactive_generation.tex
\begin{figure}
    \centering
    \includegraphics[width=\columnwidth]{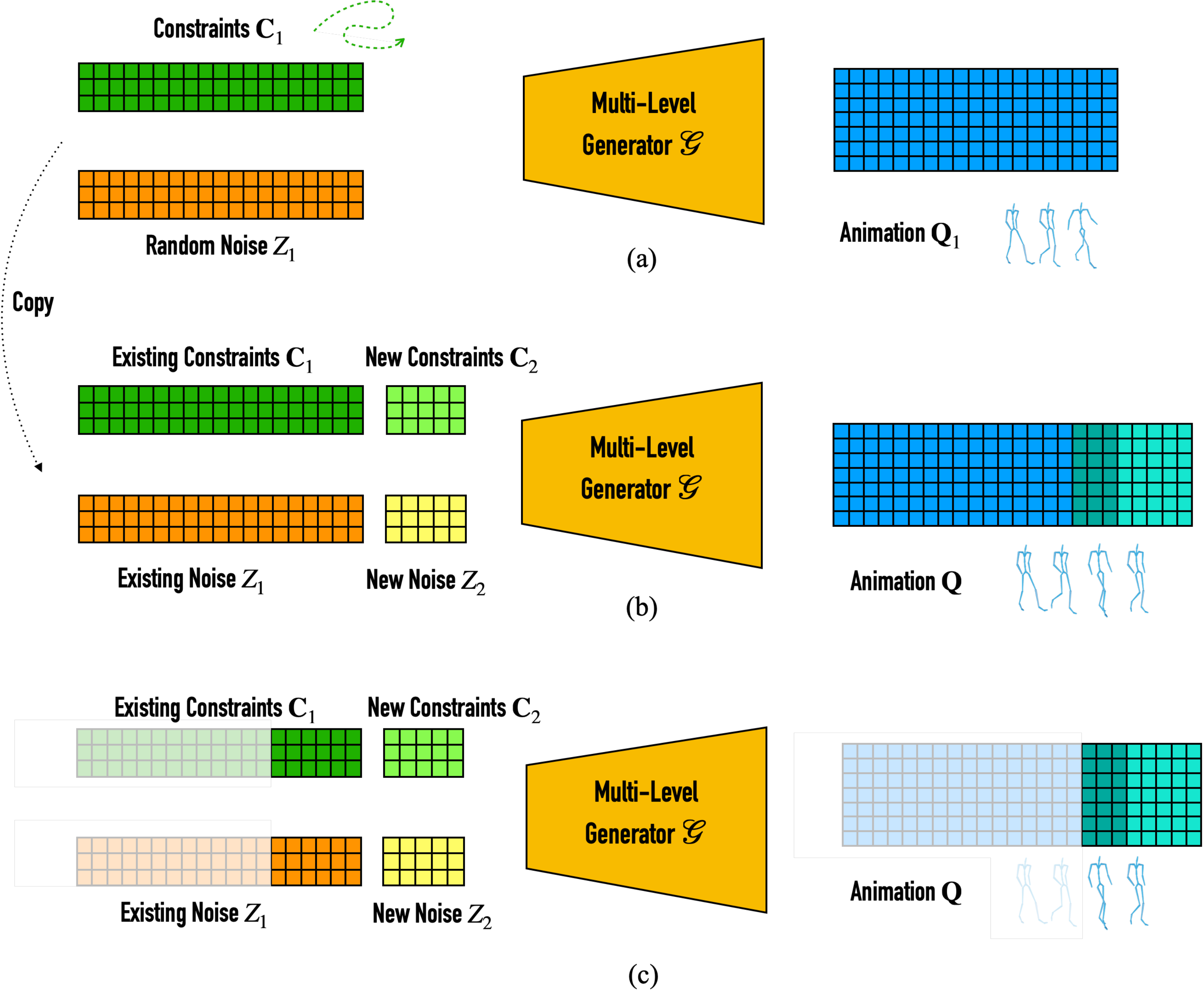}
    \caption{Interactive Generation. (a) Our conditional generation framework can be conceptually simplified as a multi-layer \emph{convolutional} generator that takes user-specified constraints e.g., root joint movement, and random noise as input to generate an animation. (b) When new constraints are given, we concatenate them with the existing constraints and noise as the input for the generator. In the generated result, the frames that are outside of the receptive field of new constraints remain the same (blue area). The frames within the receptive field of new constraints are changed and are used to create a smooth transition between existing and new constraints (dark cyan). The frames complying with new constraints are generated (cyan). (c) During a generation, we only need to keep the frames within the receptive field of the dark cyan area, denoted by activated colors.}
    \label{fig:interactive_gen}
\end{figure}

%% file: figures/016_trajectory_control.tex
\begin{figure}
    \centering
    \newcommand{\pll}{-5}
    \includegraphics[width=\columnwidth]{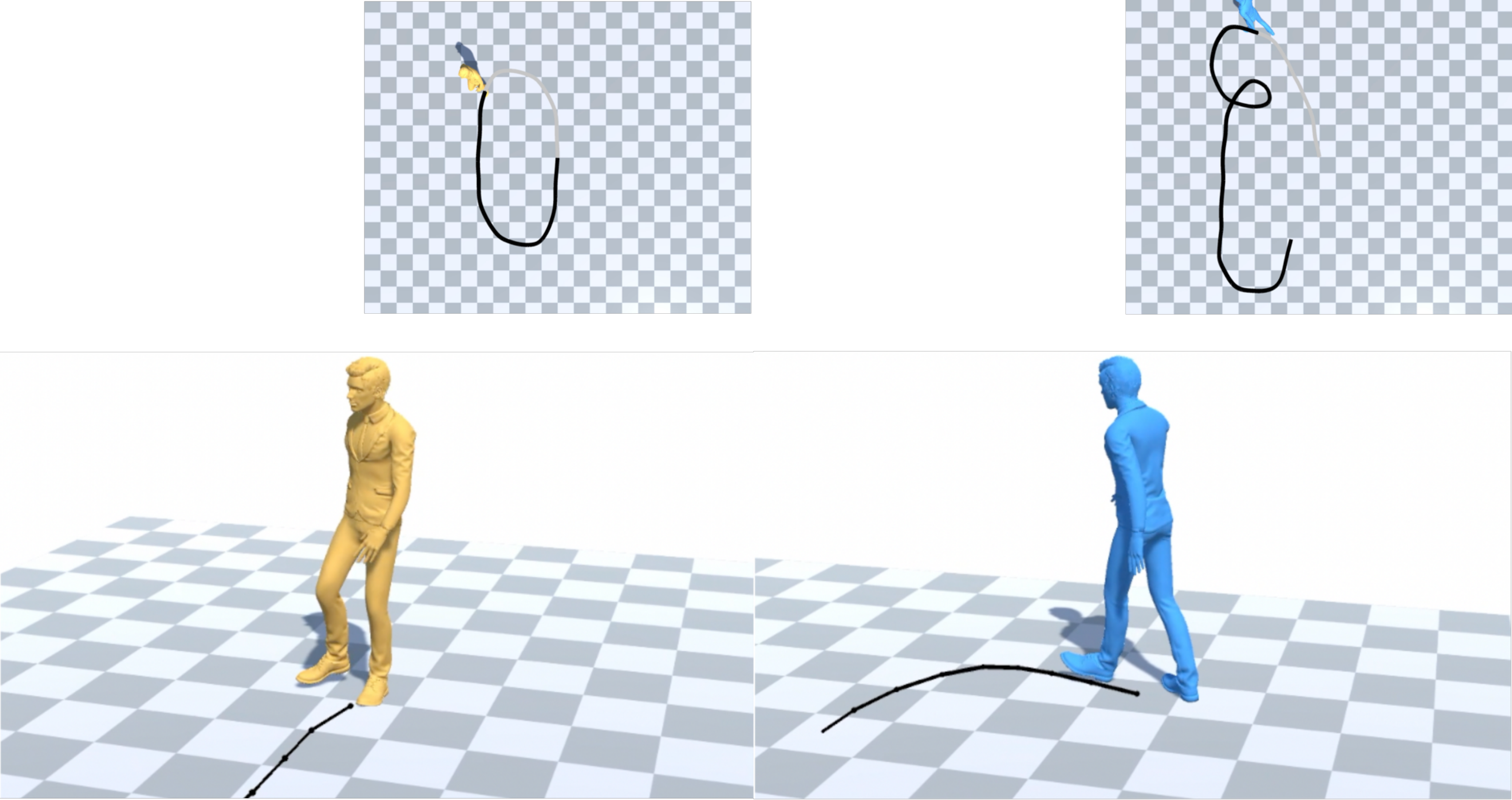}
    \begin{overpic}[width=\columnwidth]{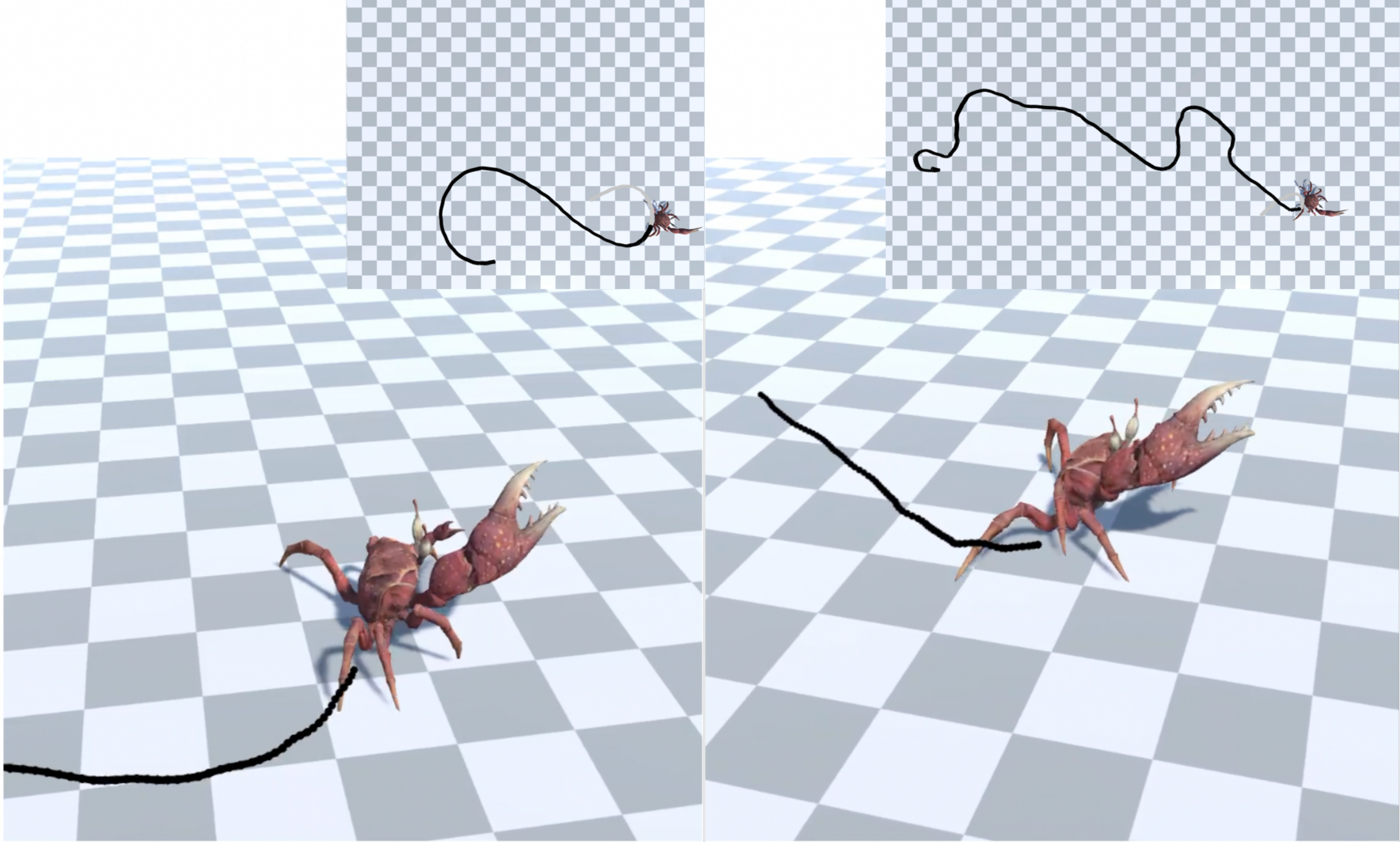}
    \put(16,  \pll){\textcolor{black}{\small Input motion}}
    \put(62,  \pll){\textcolor{black}{\small Interactive generation}}
    \end{overpic}
    \setlength{\abovecaptionskip}{10pt}
    \caption{We demonstrate the interactive trajectory control on humanoid and a hexapod crab. It can be seen that our model can cope with diverse input trajectories despite the single training sequence. }
    \label{fig:trajectory_control}
\end{figure}

%% file: conclusion.tex
\section{Discussion and Conclusion}
In this work we presented a neural motion synthesis approach that leverages the power of neural networks to exploit the information available within a \emph{single} motion sequence. We demonstrated the utility of our framework on a variety of applications, including synthesizing longer motion sequences of the same essence but with plausible variations, controlled motion synthesis, key-frame editing and interpolation, and style transfer. Despite the fact that motion data is irregular, we presented a neural representation and system for effectively learning to synthesize motion. Key to our technique is a combination of skeleton-aware convolutional operators, which serve as a backbone for a progressive motion synthesis framework.

While our current framework enables interactive control of the synthesized motion trajectory, the motion essence itself has to be learned offline, in advance. In addition, our current system's physical plausibility is limited to simple skeletal kinematics and foot contact handling. It would be interesting to incorporate higher level physics to enable synthesis of interactions and motions in complex environments.
An interesting future direction is to explore online motion learning through sparse demonstrations, for example using continual learning. It is conceivable that our method could be used for data augmentation to benefit training of elaborate frameworks that require large datasets. It is also possible to explore applications of our ideas in non-skeletal settings, such as facial and other non-skeletal rigs. Our current approach does not involve the interplay of motion and shape deformation, although the latter is also an essential part of realistic character animation (e.g., geometric ground contact for creatures like snake demonstrated in the supplementary video), and we are keen on exploring this intersection in future research.

%% file: acks.tex
\begin{acks}
    We thank Andreas Aristidou and Sigal Raab for their valuable inputs that helped to improve this work. This work is supported in part by the European Research Council (ERC) Consolidator grant no. 101003104.
\end{acks}

%% file: appendix.tex
\appendix

\section{Network Architectures}

\label{sec:netarch}

In this section, we describe the details for the network architectures.

Table~\ref{tab:arch} describes the architecture for our generator and discriminator network for a single level, where \texttt{Conv} and  \texttt{LReLU} denote skeleton-aware convolution~\cite{aberman2020skeleton} and leaky ReLU activation respectively. All the convolution layers use reflected padding, kernel size $5$ and neighbor distance $2$. For simplicity, we denote the number of input animation features $JQ + 3 + |\Jfoot|$ by $F_0$.

In our experiments, we use $\lambda_{\text{adv}} = 1, \lambda_{\text{rec}} = 50, \lambda_{\text{con}} = 5$ and $\lambda_{\text{gp}} = 1$.

\begin{table}
\begin{center}
    \begin{tabular}{ l l l l}
    \toprule
    Name &  Layers  & in/out channels\\
    \toprule Generator
     & \texttt{Conv + LReLU}  &  $F_0/F_0$                      \\
     & \texttt{Conv + LReLU}  &  $F_0/2F_0$                     \\
     & \texttt{Conv + LReLU}  &  $2F_0/2F_0$                    \\
     & \texttt{Conv}          &  $2F_0/F_0$                     \\
    \midrule Discriminator
     & \texttt{Conv + LReLU}  &  $F_0/F_0$                      \\
     & \texttt{Conv + LReLU}  &  $F_0/2F_0$                     \\
     & \texttt{Conv + LReLU}  &  $2F_0/2F_0$                    \\
     & \texttt{Conv}          &  $2F_0/1$                       \\
    \bottomrule
    \end{tabular}
\end{center}
\caption{Network Architectures}
\label{tab:arch}
\end{table}

\section{Solving Patched Nearest Neighbor}

\label{sec:dp_detail}

In this section, we describe how to solve the patched nearest neighbor (PNN). Given the generated animation $\bbq$ of length $L_Q$ and the training animation $\bbt$ of length $L_T$, we search for the corresponding segmentation $\{l_i\}_{i=1}^{L_Q}$ for each frame in $\bbq$, such that the minimum distance of any two points in the discontinuous point set $\{i \mid l_{i} \not= l_{i-1} + 1 \}$ is no less than $T_\min$. 

Let $D(i)$ denotes the PNN for first $i$ frames, with boundary condition $D(0) = 0$. We can solve other $D(i)$ by
\begin{eqnarray}
    D(i) &=& \min_{0 \leq j \leq i - T_\min, 1 \leq k \leq L_T - i + j} D(j) + \text{cost}(j, i, k) \\
    J(i), K(i) &=& \argmin_{0 \leq j \leq i - T_\min, 1 \leq k \leq L_T - i + j} D(j) + \text{cost}(j, i, k) \\
    \text{cost}(j, i, k) &=& \sum_{m=1}^{i-j}\| \bbq^{j+m} - \bbt^{k+m} \|_2^2,
\end{eqnarray}
where $\text{cost}(j, i, k)$ denotes the distance between $\bbq^{j+1:i}$ and $\bbt^{k+1:k+i-j}$. After solving $D(i), J(i), K(i)$, the label $\{l_i\}$ can be backtraced by Algorithm~\ref{alg:backtracing_label}.

\begin{algorithm}
\caption{Backtracing Label}
\label{alg:backtracing_label}
\begin{algorithmic}
    \State $p \gets L_Q$
    \While{$p > 0$}
        \State $l_{J(p)+1:p} \gets K(p) + 1, K(p) + 2, \cdots, K(p) + p - J(p)$
        \State $p \gets J(p)$
    \EndWhile
\end{algorithmic}
\end{algorithm}

The $\text{cost}(j, i, k)$ can be precomputed with time complexity $O(L_Q^2L_T)$. The dynamic programming for solving $D(i)$ also runs with time complexity $O(L_Q^2 L_T)$. The PNN cost $\Loss_{\text{PNN}}$ is given by $D(L_Q) / L_Q$.

%% file: main.bbl

\begin{thebibliography}{73}


\ifx \showCODEN    \undefined \def \showCODEN     #1{\unskip}     \fi
\ifx \showDOI      \undefined \def \showDOI       #1{#1}\fi
\ifx \showISBNx    \undefined \def \showISBNx     #1{\unskip}     \fi
\ifx \showISBNxiii \undefined \def \showISBNxiii  #1{\unskip}     \fi
\ifx \showISSN     \undefined \def \showISSN      #1{\unskip}     \fi
\ifx \showLCCN     \undefined \def \showLCCN      #1{\unskip}     \fi
\ifx \shownote     \undefined \def \shownote      #1{#1}          \fi
\ifx \showarticletitle \undefined \def \showarticletitle #1{#1}   \fi
\ifx \showURL      \undefined \def \showURL       {\relax}        \fi
\providecommand\bibfield[2]{#2}
\providecommand\bibinfo[2]{#2}
\providecommand\natexlab[1]{#1}
\providecommand\showeprint[2][]{arXiv:#2}

\bibitem[\protect\citeauthoryear{Aberman, Li, Lischinski, Sorkine-Hornung,
  Cohen-Or, and Chen}{Aberman et~al\mbox{.}}{2020a}]%
        {aberman2020skeleton}
\bibfield{author}{\bibinfo{person}{Kfir Aberman}, \bibinfo{person}{Peizhuo Li},
  \bibinfo{person}{Dani Lischinski}, \bibinfo{person}{Olga Sorkine-Hornung},
  \bibinfo{person}{Daniel Cohen-Or}, {and} \bibinfo{person}{Baoquan Chen}.}
  \bibinfo{year}{2020}\natexlab{a}.
\newblock \showarticletitle{Skeleton-aware networks for deep motion
  retargeting}.
\newblock \bibinfo{journal}{\emph{ACM Transactions on Graphics (TOG)}}
  \bibinfo{volume}{39}, \bibinfo{number}{4} (\bibinfo{year}{2020}),
  \bibinfo{pages}{62--1}.
\newblock


\bibitem[\protect\citeauthoryear{Aberman, Weng, Lischinski, Cohen-Or, and
  Chen}{Aberman et~al\mbox{.}}{2020b}]%
        {aberman2020unpaired}
\bibfield{author}{\bibinfo{person}{Kfir Aberman}, \bibinfo{person}{Yijia Weng},
  \bibinfo{person}{Dani Lischinski}, \bibinfo{person}{Daniel Cohen-Or}, {and}
  \bibinfo{person}{Baoquan Chen}.} \bibinfo{year}{2020}\natexlab{b}.
\newblock \showarticletitle{Unpaired motion style transfer from video to
  animation}.
\newblock \bibinfo{journal}{\emph{ACM Transactions on Graphics (TOG)}}
  \bibinfo{volume}{39}, \bibinfo{number}{4} (\bibinfo{year}{2020}),
  \bibinfo{pages}{64--1}.
\newblock


\bibitem[\protect\citeauthoryear{Aberman, Wu, Lischinski, Chen, and
  Cohen-Or}{Aberman et~al\mbox{.}}{2019}]%
        {aberman2019learning}
\bibfield{author}{\bibinfo{person}{Kfir Aberman}, \bibinfo{person}{Rundi Wu},
  \bibinfo{person}{Dani Lischinski}, \bibinfo{person}{Baoquan Chen}, {and}
  \bibinfo{person}{Daniel Cohen-Or}.} \bibinfo{year}{2019}\natexlab{}.
\newblock \showarticletitle{Learning Character-Agnostic Motion for Motion
  Retargeting in {2D}}.
\newblock \bibinfo{journal}{\emph{ACM Trans.~Graph.}} \bibinfo{volume}{38},
  \bibinfo{number}{4} (\bibinfo{year}{2019}), \bibinfo{pages}{75}.
\newblock


\bibitem[\protect\citeauthoryear{{Adobe Systems Inc.}}{{Adobe Systems
  Inc.}}{2021}]%
        {mixamo}
\bibfield{author}{\bibinfo{person}{{Adobe Systems Inc.}}}
  \bibinfo{year}{2021}\natexlab{}.
\newblock \bibinfo{title}{Mixamo}.
\newblock
\newblock
\urldef\tempurl%
\url{https://www.mixamo.com}
\showURL{%
\tempurl}
\newblock
\shownote{Accessed: 2021-12-25.}


\bibitem[\protect\citeauthoryear{Agrawal, Shen, and van~de Panne}{Agrawal
  et~al\mbox{.}}{2013}]%
        {agrawal2013diverse}
\bibfield{author}{\bibinfo{person}{Shailen Agrawal}, \bibinfo{person}{Shuo
  Shen}, {and} \bibinfo{person}{Michiel van~de Panne}.}
  \bibinfo{year}{2013}\natexlab{}.
\newblock \showarticletitle{Diverse motion variations for physics-based
  character animation}. In \bibinfo{booktitle}{\emph{Proceedings of the 12th
  ACM SIGGRAPH/Eurographics Symposium on Computer Animation}}.
  \bibinfo{pages}{37--44}.
\newblock


\bibitem[\protect\citeauthoryear{Arikan and Forsyth}{Arikan and
  Forsyth}{2002}]%
        {arikan2002interactive}
\bibfield{author}{\bibinfo{person}{Okan Arikan} {and} \bibinfo{person}{David~A
  Forsyth}.} \bibinfo{year}{2002}\natexlab{}.
\newblock \showarticletitle{Interactive motion generation from examples}.
\newblock \bibinfo{journal}{\emph{ACM Transactions on Graphics (TOG)}}
  \bibinfo{volume}{21}, \bibinfo{number}{3} (\bibinfo{year}{2002}),
  \bibinfo{pages}{483--490}.
\newblock


\bibitem[\protect\citeauthoryear{Aristidou, Yiannakidis, Aberman, Cohen-Or,
  Shamir, and Chrysanthou}{Aristidou et~al\mbox{.}}{2021}]%
        {aristidou2021rhythm}
\bibfield{author}{\bibinfo{person}{Andreas Aristidou},
  \bibinfo{person}{Anastasios Yiannakidis}, \bibinfo{person}{Kfir Aberman},
  \bibinfo{person}{Daniel Cohen-Or}, \bibinfo{person}{Ariel Shamir}, {and}
  \bibinfo{person}{Yiorgos Chrysanthou}.} \bibinfo{year}{2021}\natexlab{}.
\newblock \showarticletitle{Rhythm is a Dancer: Music-Driven Motion Synthesis
  with Global Structure}.
\newblock \bibinfo{journal}{\emph{arXiv preprint arXiv:2111.12159}}
  (\bibinfo{year}{2021}).
\newblock


\bibitem[\protect\citeauthoryear{Bowden}{Bowden}{2000}]%
        {bowden2000learning}
\bibfield{author}{\bibinfo{person}{Richard Bowden}.}
  \bibinfo{year}{2000}\natexlab{}.
\newblock \showarticletitle{Learning statistical models of human motion}. In
  \bibinfo{booktitle}{\emph{IEEE Workshop on Human Modeling, Analysis and
  Synthesis, CVPR}}, Vol.~\bibinfo{volume}{2000}. Citeseer.
\newblock


\bibitem[\protect\citeauthoryear{Brand and Hertzmann}{Brand and
  Hertzmann}{2000}]%
        {brand2000style}
\bibfield{author}{\bibinfo{person}{Matthew Brand} {and} \bibinfo{person}{Aaron
  Hertzmann}.} \bibinfo{year}{2000}\natexlab{}.
\newblock \showarticletitle{Style machines}. In
  \bibinfo{booktitle}{\emph{Proceedings of the 27th annual conference on
  Computer graphics and interactive techniques}}. \bibinfo{pages}{183--192}.
\newblock


\bibitem[\protect\citeauthoryear{B{\"u}ttner and Clavet}{B{\"u}ttner and
  Clavet}{2015}]%
        {buttner2015motion}
\bibfield{author}{\bibinfo{person}{Michael B{\"u}ttner} {and}
  \bibinfo{person}{Simon Clavet}.} \bibinfo{year}{2015}\natexlab{}.
\newblock \showarticletitle{Motion Matching-The Road to Next Gen Animation}.
\newblock \bibinfo{journal}{\emph{Proc. of Nucl. ai}}  \bibinfo{volume}{2015}
  (\bibinfo{year}{2015}).
\newblock
\urldef\tempurl%
\url{https://www.youtube.com/watch?v= z_wpgHFSWss&t=658s}
\showURL{%
\tempurl}


\bibitem[\protect\citeauthoryear{Chai and Hodgins}{Chai and Hodgins}{2007}]%
        {chai2007constraint}
\bibfield{author}{\bibinfo{person}{Jinxiang Chai} {and}
  \bibinfo{person}{Jessica~K Hodgins}.} \bibinfo{year}{2007}\natexlab{}.
\newblock \showarticletitle{Constraint-based motion optimization using a
  statistical dynamic model}.
\newblock In \bibinfo{booktitle}{\emph{ACM SIGGRAPH 2007 papers}}.
  \bibinfo{pages}{8--es}.
\newblock


\bibitem[\protect\citeauthoryear{Fragkiadaki, Levine, Felsen, and
  Malik}{Fragkiadaki et~al\mbox{.}}{2015}]%
        {fragkiadaki2015recurrent}
\bibfield{author}{\bibinfo{person}{Katerina Fragkiadaki},
  \bibinfo{person}{Sergey Levine}, \bibinfo{person}{Panna Felsen}, {and}
  \bibinfo{person}{Jitendra Malik}.} \bibinfo{year}{2015}\natexlab{}.
\newblock \showarticletitle{Recurrent network models for human dynamics}. In
  \bibinfo{booktitle}{\emph{Proceedings of the IEEE International Conference on
  Computer Vision}}. \bibinfo{pages}{4346--4354}.
\newblock


\bibitem[\protect\citeauthoryear{Geijtenbeek, Pronost, Egges, and
  Overmars}{Geijtenbeek et~al\mbox{.}}{2011}]%
        {PhysicsCharacterAnimationSTAR:2011}
\bibfield{author}{\bibinfo{person}{Thomas Geijtenbeek},
  \bibinfo{person}{Nicolas Pronost}, \bibinfo{person}{Arjan Egges}, {and}
  \bibinfo{person}{Mark~H. Overmars}.} \bibinfo{year}{2011}\natexlab{}.
\newblock \showarticletitle{Interactive Character Animation using Simulated
  Physics}. In \bibinfo{booktitle}{\emph{Eurographics 2011 - State of the Art
  Reports}}, \bibfield{editor}{\bibinfo{person}{N.\ John} {and}
  \bibinfo{person}{B.\ Wyvill}} (Eds.). \bibinfo{publisher}{The Eurographics
  Association}.
\newblock
\showISSN{1017-4656}
\urldef\tempurl%
\url{https://doi.org/10.2312/EG2011/stars/127-149}
\showDOI{\tempurl}


\bibitem[\protect\citeauthoryear{Goodfellow, Pouget-Abadie, Mirza, Xu,
  Warde-Farley, Ozair, Courville, and Bengio}{Goodfellow et~al\mbox{.}}{2014}]%
        {goodfellow2014generative}
\bibfield{author}{\bibinfo{person}{Ian Goodfellow}, \bibinfo{person}{Jean
  Pouget-Abadie}, \bibinfo{person}{Mehdi Mirza}, \bibinfo{person}{Bing Xu},
  \bibinfo{person}{David Warde-Farley}, \bibinfo{person}{Sherjil Ozair},
  \bibinfo{person}{Aaron Courville}, {and} \bibinfo{person}{Yoshua Bengio}.}
  \bibinfo{year}{2014}\natexlab{}.
\newblock \showarticletitle{Generative adversarial nets}.
\newblock \bibinfo{journal}{\emph{Advances in neural information processing
  systems}}  \bibinfo{volume}{27} (\bibinfo{year}{2014}).
\newblock


\bibitem[\protect\citeauthoryear{Grochow, Martin, Hertzmann, and
  Popovi{\'c}}{Grochow et~al\mbox{.}}{2004}]%
        {grochow2004style}
\bibfield{author}{\bibinfo{person}{Keith Grochow}, \bibinfo{person}{Steven~L
  Martin}, \bibinfo{person}{Aaron Hertzmann}, {and} \bibinfo{person}{Zoran
  Popovi{\'c}}.} \bibinfo{year}{2004}\natexlab{}.
\newblock \showarticletitle{Style-based inverse kinematics}.
\newblock In \bibinfo{booktitle}{\emph{ACM SIGGRAPH 2004 Papers}}.
  \bibinfo{pages}{522--531}.
\newblock


\bibitem[\protect\citeauthoryear{Gulrajani, Ahmed, Arjovsky, Dumoulin, and
  Courville}{Gulrajani et~al\mbox{.}}{2017}]%
        {gulrajani2017improved}
\bibfield{author}{\bibinfo{person}{Ishaan Gulrajani}, \bibinfo{person}{Faruk
  Ahmed}, \bibinfo{person}{Martin Arjovsky}, \bibinfo{person}{Vincent
  Dumoulin}, {and} \bibinfo{person}{Aaron Courville}.}
  \bibinfo{year}{2017}\natexlab{}.
\newblock \showarticletitle{Improved training of wasserstein GANs}. In
  \bibinfo{booktitle}{\emph{Proceedings of the 31st International Conference on
  Neural Information Processing Systems}}. \bibinfo{pages}{5769--5779}.
\newblock


\bibitem[\protect\citeauthoryear{Harvey, Yurick, Nowrouzezahrai, and
  Pal}{Harvey et~al\mbox{.}}{2020}]%
        {harvey2020robust}
\bibfield{author}{\bibinfo{person}{F{\'e}lix~G Harvey}, \bibinfo{person}{Mike
  Yurick}, \bibinfo{person}{Derek Nowrouzezahrai}, {and}
  \bibinfo{person}{Christopher Pal}.} \bibinfo{year}{2020}\natexlab{}.
\newblock \showarticletitle{Robust motion in-betweening}.
\newblock \bibinfo{journal}{\emph{ACM Transactions on Graphics (TOG)}}
  \bibinfo{volume}{39}, \bibinfo{number}{4} (\bibinfo{year}{2020}),
  \bibinfo{pages}{60--1}.
\newblock


\bibitem[\protect\citeauthoryear{He, Zhang, Ren, and Sun}{He
  et~al\mbox{.}}{2016}]%
        {he2016deep}
\bibfield{author}{\bibinfo{person}{Kaiming He}, \bibinfo{person}{Xiangyu
  Zhang}, \bibinfo{person}{Shaoqing Ren}, {and} \bibinfo{person}{Jian Sun}.}
  \bibinfo{year}{2016}\natexlab{}.
\newblock \showarticletitle{Deep residual learning for image recognition}. In
  \bibinfo{booktitle}{\emph{Proceedings of the IEEE conference on computer
  vision and pattern recognition}}. \bibinfo{pages}{770--778}.
\newblock


\bibitem[\protect\citeauthoryear{Heck and Gleicher}{Heck and Gleicher}{2007}]%
        {heck2007parametric}
\bibfield{author}{\bibinfo{person}{Rachel Heck} {and} \bibinfo{person}{Michael
  Gleicher}.} \bibinfo{year}{2007}\natexlab{}.
\newblock \showarticletitle{Parametric motion graphs}. In
  \bibinfo{booktitle}{\emph{Proceedings of the 2007 symposium on Interactive 3D
  graphics and games}}. \bibinfo{pages}{129--136}.
\newblock


\bibitem[\protect\citeauthoryear{Heess, TB, Sriram, Lemmon, Merel, Wayne,
  Tassa, Erez, Wang, Eslami, et~al\mbox{.}}{Heess et~al\mbox{.}}{2017}]%
        {heess2017emergence}
\bibfield{author}{\bibinfo{person}{Nicolas Heess}, \bibinfo{person}{Dhruva TB},
  \bibinfo{person}{Srinivasan Sriram}, \bibinfo{person}{Jay Lemmon},
  \bibinfo{person}{Josh Merel}, \bibinfo{person}{Greg Wayne},
  \bibinfo{person}{Yuval Tassa}, \bibinfo{person}{Tom Erez},
  \bibinfo{person}{Ziyu Wang}, \bibinfo{person}{SM Eslami}, {et~al\mbox{.}}}
  \bibinfo{year}{2017}\natexlab{}.
\newblock \showarticletitle{Emergence of locomotion behaviours in rich
  environments}.
\newblock \bibinfo{journal}{\emph{arXiv preprint arXiv:1707.02286}}
  (\bibinfo{year}{2017}).
\newblock


\bibitem[\protect\citeauthoryear{Henter, Alexanderson, and Beskow}{Henter
  et~al\mbox{.}}{2020}]%
        {henter2020moglow}
\bibfield{author}{\bibinfo{person}{Gustav~Eje Henter}, \bibinfo{person}{Simon
  Alexanderson}, {and} \bibinfo{person}{Jonas Beskow}.}
  \bibinfo{year}{2020}\natexlab{}.
\newblock \showarticletitle{Moglow: Probabilistic and controllable motion
  synthesis using normalising flows}.
\newblock \bibinfo{journal}{\emph{ACM Transactions on Graphics (TOG)}}
  \bibinfo{volume}{39}, \bibinfo{number}{6} (\bibinfo{year}{2020}),
  \bibinfo{pages}{1--14}.
\newblock


\bibitem[\protect\citeauthoryear{Hinz, Fisher, Wang, and Wermter}{Hinz
  et~al\mbox{.}}{2021}]%
        {hinz2021improved}
\bibfield{author}{\bibinfo{person}{Tobias Hinz}, \bibinfo{person}{Matthew
  Fisher}, \bibinfo{person}{Oliver Wang}, {and} \bibinfo{person}{Stefan
  Wermter}.} \bibinfo{year}{2021}\natexlab{}.
\newblock \showarticletitle{Improved techniques for training single-image
  gans}. In \bibinfo{booktitle}{\emph{Proceedings of the IEEE/CVF Winter
  Conference on Applications of Computer Vision}}. \bibinfo{pages}{1300--1309}.
\newblock


\bibitem[\protect\citeauthoryear{Holden, Kanoun, Perepichka, and Popa}{Holden
  et~al\mbox{.}}{2020}]%
        {holden2020learned}
\bibfield{author}{\bibinfo{person}{Daniel Holden}, \bibinfo{person}{Oussama
  Kanoun}, \bibinfo{person}{Maksym Perepichka}, {and} \bibinfo{person}{Tiberiu
  Popa}.} \bibinfo{year}{2020}\natexlab{}.
\newblock \showarticletitle{Learned motion matching}.
\newblock \bibinfo{journal}{\emph{ACM Transactions on Graphics (TOG)}}
  \bibinfo{volume}{39}, \bibinfo{number}{4} (\bibinfo{year}{2020}),
  \bibinfo{pages}{53--1}.
\newblock


\bibitem[\protect\citeauthoryear{Holden, Komura, and Saito}{Holden
  et~al\mbox{.}}{2017}]%
        {holden2017phase}
\bibfield{author}{\bibinfo{person}{Daniel Holden}, \bibinfo{person}{Taku
  Komura}, {and} \bibinfo{person}{Jun Saito}.} \bibinfo{year}{2017}\natexlab{}.
\newblock \showarticletitle{Phase-functioned neural networks for character
  control}.
\newblock \bibinfo{journal}{\emph{ACM Transactions on Graphics (TOG)}}
  \bibinfo{volume}{36}, \bibinfo{number}{4} (\bibinfo{year}{2017}),
  \bibinfo{pages}{1--13}.
\newblock


\bibitem[\protect\citeauthoryear{Holden, Saito, and Komura}{Holden
  et~al\mbox{.}}{2016}]%
        {holden2016deep}
\bibfield{author}{\bibinfo{person}{Daniel Holden}, \bibinfo{person}{Jun Saito},
  {and} \bibinfo{person}{Taku Komura}.} \bibinfo{year}{2016}\natexlab{}.
\newblock \showarticletitle{A deep learning framework for character motion
  synthesis and editing}.
\newblock \bibinfo{journal}{\emph{ACM Transactions on Graphics (TOG)}}
  \bibinfo{volume}{35}, \bibinfo{number}{4} (\bibinfo{year}{2016}),
  \bibinfo{pages}{1--11}.
\newblock


\bibitem[\protect\citeauthoryear{Holden, Saito, Komura, and Joyce}{Holden
  et~al\mbox{.}}{2015}]%
        {holden2015learning}
\bibfield{author}{\bibinfo{person}{Daniel Holden}, \bibinfo{person}{Jun Saito},
  \bibinfo{person}{Taku Komura}, {and} \bibinfo{person}{Thomas Joyce}.}
  \bibinfo{year}{2015}\natexlab{}.
\newblock \showarticletitle{Learning motion manifolds with convolutional
  autoencoders}.
\newblock In \bibinfo{booktitle}{\emph{SIGGRAPH Asia 2015 Technical Briefs}}.
  \bibinfo{pages}{1--4}.
\newblock


\bibitem[\protect\citeauthoryear{Ikemoto, Arikan, and Forsyth}{Ikemoto
  et~al\mbox{.}}{2009}]%
        {ikemoto2009generalizing}
\bibfield{author}{\bibinfo{person}{Leslie Ikemoto}, \bibinfo{person}{Okan
  Arikan}, {and} \bibinfo{person}{David Forsyth}.}
  \bibinfo{year}{2009}\natexlab{}.
\newblock \showarticletitle{Generalizing motion edits with gaussian processes}.
\newblock \bibinfo{journal}{\emph{ACM Transactions on Graphics (TOG)}}
  \bibinfo{volume}{28}, \bibinfo{number}{1} (\bibinfo{year}{2009}),
  \bibinfo{pages}{1--12}.
\newblock


\bibitem[\protect\citeauthoryear{Isola, Zhu, Zhou, and Efros}{Isola
  et~al\mbox{.}}{2017}]%
        {isola2017image}
\bibfield{author}{\bibinfo{person}{Phillip Isola}, \bibinfo{person}{Jun-Yan
  Zhu}, \bibinfo{person}{Tinghui Zhou}, {and} \bibinfo{person}{Alexei~A
  Efros}.} \bibinfo{year}{2017}\natexlab{}.
\newblock \showarticletitle{Image-to-image translation with conditional
  adversarial networks}. In \bibinfo{booktitle}{\emph{Proceedings of the IEEE
  conference on computer vision and pattern recognition}}.
  \bibinfo{pages}{1125--1134}.
\newblock


\bibitem[\protect\citeauthoryear{Karras, Aila, Laine, and Lehtinen}{Karras
  et~al\mbox{.}}{2018}]%
        {karras2018progressive}
\bibfield{author}{\bibinfo{person}{Tero Karras}, \bibinfo{person}{Timo Aila},
  \bibinfo{person}{Samuli Laine}, {and} \bibinfo{person}{Jaakko Lehtinen}.}
  \bibinfo{year}{2018}\natexlab{}.
\newblock \showarticletitle{Progressive Growing of GANs for Improved Quality,
  Stability, and Variation}. In \bibinfo{booktitle}{\emph{International
  Conference on Learning Representations}}.
\newblock


\bibitem[\protect\citeauthoryear{Kingma and Ba}{Kingma and Ba}{2014}]%
        {kingma2014adam}
\bibfield{author}{\bibinfo{person}{Diederik~P Kingma} {and}
  \bibinfo{person}{Jimmy Ba}.} \bibinfo{year}{2014}\natexlab{}.
\newblock \showarticletitle{Adam: A method for stochastic optimization}.
\newblock \bibinfo{journal}{\emph{arXiv preprint arXiv:1412.6980}}
  (\bibinfo{year}{2014}).
\newblock


\bibitem[\protect\citeauthoryear{Kovar and Gleicher}{Kovar and
  Gleicher}{2004}]%
        {kovar2004automated}
\bibfield{author}{\bibinfo{person}{Lucas Kovar} {and} \bibinfo{person}{Michael
  Gleicher}.} \bibinfo{year}{2004}\natexlab{}.
\newblock \showarticletitle{Automated extraction and parameterization of
  motions in large data sets}.
\newblock \bibinfo{journal}{\emph{ACM Transactions on Graphics (ToG)}}
  \bibinfo{volume}{23}, \bibinfo{number}{3} (\bibinfo{year}{2004}),
  \bibinfo{pages}{559--568}.
\newblock


\bibitem[\protect\citeauthoryear{Kovar, Gleicher, and Pighin}{Kovar
  et~al\mbox{.}}{2002}]%
        {kovar2002motion}
\bibfield{author}{\bibinfo{person}{Lucas Kovar}, \bibinfo{person}{Michael
  Gleicher}, {and} \bibinfo{person}{Fr\'{e}d\'{e}ric Pighin}.}
  \bibinfo{year}{2002}\natexlab{}.
\newblock \showarticletitle{Motion Graphs}. In
  \bibinfo{booktitle}{\emph{Proceedings of the 29th Annual Conference on
  Computer Graphics and Interactive Techniques}} (San Antonio, Texas)
  \emph{(\bibinfo{series}{SIGGRAPH '02})}. \bibinfo{publisher}{Association for
  Computing Machinery}, \bibinfo{address}{New York, NY, USA},
  \bibinfo{pages}{473–482}.
\newblock
\showISBNx{1581135211}
\urldef\tempurl%
\url{https://doi.org/10.1145/566570.566605}
\showDOI{\tempurl}


\bibitem[\protect\citeauthoryear{Lau, Bar-Joseph, and Kuffner}{Lau
  et~al\mbox{.}}{2009}]%
        {lau2009modeling}
\bibfield{author}{\bibinfo{person}{Manfred Lau}, \bibinfo{person}{Ziv
  Bar-Joseph}, {and} \bibinfo{person}{James Kuffner}.}
  \bibinfo{year}{2009}\natexlab{}.
\newblock \showarticletitle{Modeling spatial and temporal variation in motion
  data}.
\newblock \bibinfo{journal}{\emph{ACM Transactions on Graphics (TOG)}}
  \bibinfo{volume}{28}, \bibinfo{number}{5} (\bibinfo{year}{2009}),
  \bibinfo{pages}{1--10}.
\newblock


\bibitem[\protect\citeauthoryear{Lee, Chai, Reitsma, Hodgins, and Pollard}{Lee
  et~al\mbox{.}}{2002}]%
        {lee2002interactive}
\bibfield{author}{\bibinfo{person}{Jehee Lee}, \bibinfo{person}{Jinxiang Chai},
  \bibinfo{person}{Paul~SA Reitsma}, \bibinfo{person}{Jessica~K Hodgins}, {and}
  \bibinfo{person}{Nancy~S Pollard}.} \bibinfo{year}{2002}\natexlab{}.
\newblock \showarticletitle{Interactive control of avatars animated with human
  motion data}. In \bibinfo{booktitle}{\emph{Proceedings of the 29th annual
  conference on Computer graphics and interactive techniques}}.
  \bibinfo{pages}{491--500}.
\newblock


\bibitem[\protect\citeauthoryear{Lee, Lee, and Lee}{Lee et~al\mbox{.}}{2018}]%
        {lee2018interactive}
\bibfield{author}{\bibinfo{person}{Kyungho Lee}, \bibinfo{person}{Seyoung Lee},
  {and} \bibinfo{person}{Jehee Lee}.} \bibinfo{year}{2018}\natexlab{}.
\newblock \showarticletitle{Interactive character animation by learning
  multi-objective control}.
\newblock \bibinfo{journal}{\emph{ACM Transactions on Graphics (TOG)}}
  \bibinfo{volume}{37}, \bibinfo{number}{6} (\bibinfo{year}{2018}),
  \bibinfo{pages}{1--10}.
\newblock


\bibitem[\protect\citeauthoryear{Lee, Lee, Lee, and Lee}{Lee
  et~al\mbox{.}}{2021}]%
        {lee2021learning}
\bibfield{author}{\bibinfo{person}{Seyoung Lee}, \bibinfo{person}{Sunmin Lee},
  \bibinfo{person}{Yongwoo Lee}, {and} \bibinfo{person}{Jehee Lee}.}
  \bibinfo{year}{2021}\natexlab{}.
\newblock \showarticletitle{Learning a family of motor skills from a single
  motion clip}.
\newblock \bibinfo{journal}{\emph{ACM Transactions on Graphics (TOG)}}
  \bibinfo{volume}{40}, \bibinfo{number}{4} (\bibinfo{year}{2021}),
  \bibinfo{pages}{1--13}.
\newblock


\bibitem[\protect\citeauthoryear{Levine, Wang, Haraux, Popovi{\'c}, and
  Koltun}{Levine et~al\mbox{.}}{2012}]%
        {levine2012continuous}
\bibfield{author}{\bibinfo{person}{Sergey Levine}, \bibinfo{person}{Jack~M
  Wang}, \bibinfo{person}{Alexis Haraux}, \bibinfo{person}{Zoran Popovi{\'c}},
  {and} \bibinfo{person}{Vladlen Koltun}.} \bibinfo{year}{2012}\natexlab{}.
\newblock \showarticletitle{Continuous character control with low-dimensional
  embeddings}.
\newblock \bibinfo{journal}{\emph{ACM Transactions on Graphics (TOG)}}
  \bibinfo{volume}{31}, \bibinfo{number}{4} (\bibinfo{year}{2012}),
  \bibinfo{pages}{1--10}.
\newblock


\bibitem[\protect\citeauthoryear{Li and Wand}{Li and Wand}{2016}]%
        {li2016precomputed}
\bibfield{author}{\bibinfo{person}{Chuan Li} {and} \bibinfo{person}{Michael
  Wand}.} \bibinfo{year}{2016}\natexlab{}.
\newblock \showarticletitle{Precomputed real-time texture synthesis with
  markovian generative adversarial networks}. In
  \bibinfo{booktitle}{\emph{European conference on computer vision}}. Springer,
  \bibinfo{pages}{702--716}.
\newblock


\bibitem[\protect\citeauthoryear{Li, Wang, and Shum}{Li et~al\mbox{.}}{2002}]%
        {li2002motion}
\bibfield{author}{\bibinfo{person}{Yan Li}, \bibinfo{person}{Tianshu Wang},
  {and} \bibinfo{person}{Heung-Yeung Shum}.} \bibinfo{year}{2002}\natexlab{}.
\newblock \showarticletitle{Motion texture: a two-level statistical model for
  character motion synthesis}. In \bibinfo{booktitle}{\emph{Proceedings of the
  29th annual conference on Computer graphics and interactive techniques}}.
  \bibinfo{pages}{465--472}.
\newblock


\bibitem[\protect\citeauthoryear{Luo, Soeseno, Chen, and Chen}{Luo
  et~al\mbox{.}}{2020}]%
        {luo2020carl}
\bibfield{author}{\bibinfo{person}{Ying-Sheng Luo},
  \bibinfo{person}{Jonathan~Hans Soeseno}, \bibinfo{person}{Trista Pei-Chun
  Chen}, {and} \bibinfo{person}{Wei-Chao Chen}.}
  \bibinfo{year}{2020}\natexlab{}.
\newblock \showarticletitle{Carl: Controllable agent with reinforcement
  learning for quadruped locomotion}.
\newblock \bibinfo{journal}{\emph{ACM Transactions on Graphics (TOG)}}
  \bibinfo{volume}{39}, \bibinfo{number}{4} (\bibinfo{year}{2020}),
  \bibinfo{pages}{38--1}.
\newblock


\bibitem[\protect\citeauthoryear{Mason, Starke, and Komura}{Mason
  et~al\mbox{.}}{2022}]%
        {mason2022real}
\bibfield{author}{\bibinfo{person}{Ian Mason}, \bibinfo{person}{Sebastian
  Starke}, {and} \bibinfo{person}{Taku Komura}.}
  \bibinfo{year}{2022}\natexlab{}.
\newblock \showarticletitle{Real-Time Style Modelling of Human Locomotion via
  Feature-Wise Transformations and Local Motion Phases}.
\newblock \bibinfo{journal}{\emph{arXiv preprint arXiv:2201.04439}}
  (\bibinfo{year}{2022}).
\newblock


\bibitem[\protect\citeauthoryear{Min and Chai}{Min and Chai}{2012}]%
        {min2012motion}
\bibfield{author}{\bibinfo{person}{Jianyuan Min} {and}
  \bibinfo{person}{Jinxiang Chai}.} \bibinfo{year}{2012}\natexlab{}.
\newblock \showarticletitle{Motion graphs++ a compact generative model for
  semantic motion analysis and synthesis}.
\newblock \bibinfo{journal}{\emph{ACM Transactions on Graphics (TOG)}}
  \bibinfo{volume}{31}, \bibinfo{number}{6} (\bibinfo{year}{2012}),
  \bibinfo{pages}{1--12}.
\newblock


\bibitem[\protect\citeauthoryear{Mizuguchi, Buchanan, and Calvert}{Mizuguchi
  et~al\mbox{.}}{2001}]%
        {mizuguchi2001data}
\bibfield{author}{\bibinfo{person}{Mark Mizuguchi}, \bibinfo{person}{John
  Buchanan}, {and} \bibinfo{person}{Tom Calvert}.}
  \bibinfo{year}{2001}\natexlab{}.
\newblock \showarticletitle{Data driven motion transitions for interactive
  games.}. In \bibinfo{booktitle}{\emph{Eurographics (Short Presentations)}}.
\newblock


\bibitem[\protect\citeauthoryear{Mourot, Hoyet, Le~Clerc, Schnitzler, and
  Hellier}{Mourot et~al\mbox{.}}{2021}]%
        {mourot2021survey}
\bibfield{author}{\bibinfo{person}{Lucas Mourot}, \bibinfo{person}{Ludovic
  Hoyet}, \bibinfo{person}{Fran{\c{c}}ois Le~Clerc},
  \bibinfo{person}{Fran{\c{c}}ois Schnitzler}, {and} \bibinfo{person}{Pierre
  Hellier}.} \bibinfo{year}{2021}\natexlab{}.
\newblock \showarticletitle{A Survey on Deep Learning for Skeleton-Based Human
  Animation}.
\newblock \bibinfo{journal}{\emph{Computer Graphics Forum}}
  (\bibinfo{year}{2021}).
\newblock


\bibitem[\protect\citeauthoryear{Oord, Dieleman, Zen, Simonyan, Vinyals,
  Graves, Kalchbrenner, Senior, and Kavukcuoglu}{Oord et~al\mbox{.}}{2016}]%
        {oord2016wavenet}
\bibfield{author}{\bibinfo{person}{Aaron van~den Oord}, \bibinfo{person}{Sander
  Dieleman}, \bibinfo{person}{Heiga Zen}, \bibinfo{person}{Karen Simonyan},
  \bibinfo{person}{Oriol Vinyals}, \bibinfo{person}{Alex Graves},
  \bibinfo{person}{Nal Kalchbrenner}, \bibinfo{person}{Andrew Senior}, {and}
  \bibinfo{person}{Koray Kavukcuoglu}.} \bibinfo{year}{2016}\natexlab{}.
\newblock \showarticletitle{Wavenet: A generative model for raw audio}.
\newblock \bibinfo{journal}{\emph{arXiv preprint arXiv:1609.03499}}
  (\bibinfo{year}{2016}).
\newblock


\bibitem[\protect\citeauthoryear{Park, Shin, and Shin}{Park
  et~al\mbox{.}}{2002}]%
        {park2002line}
\bibfield{author}{\bibinfo{person}{Sang~Il Park}, \bibinfo{person}{Hyun~Joon
  Shin}, {and} \bibinfo{person}{Sung~Yong Shin}.}
  \bibinfo{year}{2002}\natexlab{}.
\newblock \showarticletitle{On-line locomotion generation based on motion
  blending}. In \bibinfo{booktitle}{\emph{Proceedings of the 2002 ACM
  SIGGRAPH/Eurographics symposium on Computer animation}}.
  \bibinfo{pages}{105--111}.
\newblock


\bibitem[\protect\citeauthoryear{Paszke, Gross, Massa, Lerer, Bradbury, Chanan,
  Killeen, Lin, Gimelshein, Antiga, Desmaison, Kopf, Yang, DeVito, Raison,
  Tejani, Chilamkurthy, Steiner, Fang, Bai, and Chintala}{Paszke
  et~al\mbox{.}}{2019}]%
        {NEURIPS2019_9015}
\bibfield{author}{\bibinfo{person}{Adam Paszke}, \bibinfo{person}{Sam Gross},
  \bibinfo{person}{Francisco Massa}, \bibinfo{person}{Adam Lerer},
  \bibinfo{person}{James Bradbury}, \bibinfo{person}{Gregory Chanan},
  \bibinfo{person}{Trevor Killeen}, \bibinfo{person}{Zeming Lin},
  \bibinfo{person}{Natalia Gimelshein}, \bibinfo{person}{Luca Antiga},
  \bibinfo{person}{Alban Desmaison}, \bibinfo{person}{Andreas Kopf},
  \bibinfo{person}{Edward Yang}, \bibinfo{person}{Zachary DeVito},
  \bibinfo{person}{Martin Raison}, \bibinfo{person}{Alykhan Tejani},
  \bibinfo{person}{Sasank Chilamkurthy}, \bibinfo{person}{Benoit Steiner},
  \bibinfo{person}{Lu Fang}, \bibinfo{person}{Junjie Bai}, {and}
  \bibinfo{person}{Soumith Chintala}.} \bibinfo{year}{2019}\natexlab{}.
\newblock \showarticletitle{PyTorch: An Imperative Style, High-Performance Deep
  Learning Library}.
\newblock In \bibinfo{booktitle}{\emph{Advances in Neural Information
  Processing Systems 32}}, \bibfield{editor}{\bibinfo{person}{H.~Wallach},
  \bibinfo{person}{H.~Larochelle}, \bibinfo{person}{A.~Beygelzimer},
  \bibinfo{person}{F.~d\textquotesingle Alch\'{e}-Buc},
  \bibinfo{person}{E.~Fox}, {and} \bibinfo{person}{R.~Garnett}} (Eds.).
  \bibinfo{publisher}{Curran Associates, Inc.}, \bibinfo{pages}{8024--8035}.
\newblock
\urldef\tempurl%
\url{http://papers.neurips.cc/paper/9015-pytorch-an-imperative-style-high-performance-deep-learning-library.pdf}
\showURL{%
\tempurl}


\bibitem[\protect\citeauthoryear{Pavllo, Grangier, and Auli}{Pavllo
  et~al\mbox{.}}{2018}]%
        {pavllo2018quaternet}
\bibfield{author}{\bibinfo{person}{Dario Pavllo}, \bibinfo{person}{David
  Grangier}, {and} \bibinfo{person}{Michael Auli}.}
  \bibinfo{year}{2018}\natexlab{}.
\newblock \showarticletitle{Quaternet: A quaternion-based recurrent model for
  human motion}.
\newblock \bibinfo{journal}{\emph{arXiv preprint arXiv:1805.06485}}
  (\bibinfo{year}{2018}).
\newblock


\bibitem[\protect\citeauthoryear{Peng, Abbeel, Levine, and van~de Panne}{Peng
  et~al\mbox{.}}{2018}]%
        {peng2018deepmimic}
\bibfield{author}{\bibinfo{person}{Xue~Bin Peng}, \bibinfo{person}{Pieter
  Abbeel}, \bibinfo{person}{Sergey Levine}, {and} \bibinfo{person}{Michiel
  van~de Panne}.} \bibinfo{year}{2018}\natexlab{}.
\newblock \showarticletitle{Deepmimic: Example-guided deep reinforcement
  learning of physics-based character skills}.
\newblock \bibinfo{journal}{\emph{ACM Transactions on Graphics (TOG)}}
  \bibinfo{volume}{37}, \bibinfo{number}{4} (\bibinfo{year}{2018}),
  \bibinfo{pages}{1--14}.
\newblock


\bibitem[\protect\citeauthoryear{Peng, Berseth, Yin, and Van De~Panne}{Peng
  et~al\mbox{.}}{2017}]%
        {peng2017deeploco}
\bibfield{author}{\bibinfo{person}{Xue~Bin Peng}, \bibinfo{person}{Glen
  Berseth}, \bibinfo{person}{KangKang Yin}, {and} \bibinfo{person}{Michiel Van
  De~Panne}.} \bibinfo{year}{2017}\natexlab{}.
\newblock \showarticletitle{Deeploco: Dynamic locomotion skills using
  hierarchical deep reinforcement learning}.
\newblock \bibinfo{journal}{\emph{ACM Transactions on Graphics (TOG)}}
  \bibinfo{volume}{36}, \bibinfo{number}{4} (\bibinfo{year}{2017}),
  \bibinfo{pages}{1--13}.
\newblock


\bibitem[\protect\citeauthoryear{Perlin}{Perlin}{1985}]%
        {perlin1985image}
\bibfield{author}{\bibinfo{person}{Ken Perlin}.}
  \bibinfo{year}{1985}\natexlab{}.
\newblock \showarticletitle{An image synthesizer}.
\newblock \bibinfo{journal}{\emph{ACM Siggraph Computer Graphics}}
  \bibinfo{volume}{19}, \bibinfo{number}{3} (\bibinfo{year}{1985}),
  \bibinfo{pages}{287--296}.
\newblock


\bibitem[\protect\citeauthoryear{Perlin and Goldberg}{Perlin and
  Goldberg}{1996}]%
        {perlin1996improv}
\bibfield{author}{\bibinfo{person}{Ken Perlin} {and} \bibinfo{person}{Athomas
  Goldberg}.} \bibinfo{year}{1996}\natexlab{}.
\newblock \showarticletitle{Improv: A system for scripting interactive actors
  in virtual worlds}. In \bibinfo{booktitle}{\emph{Proceedings of the 23rd
  annual conference on Computer graphics and interactive techniques}}.
  \bibinfo{pages}{205--216}.
\newblock


\bibitem[\protect\citeauthoryear{Pullen and Bregler}{Pullen and
  Bregler}{2000}]%
        {pullen2000animating}
\bibfield{author}{\bibinfo{person}{Katherine Pullen} {and}
  \bibinfo{person}{Christoph Bregler}.} \bibinfo{year}{2000}\natexlab{}.
\newblock \showarticletitle{Animating by multi-level sampling}. In
  \bibinfo{booktitle}{\emph{Proceedings Computer Animation 2000}}. IEEE,
  \bibinfo{pages}{36--42}.
\newblock


\bibitem[\protect\citeauthoryear{Pullen and Bregler}{Pullen and
  Bregler}{2002}]%
        {pullen2002motion}
\bibfield{author}{\bibinfo{person}{Katherine Pullen} {and}
  \bibinfo{person}{Christoph Bregler}.} \bibinfo{year}{2002}\natexlab{}.
\newblock \showarticletitle{Motion capture assisted animation: Texturing and
  synthesis}. In \bibinfo{booktitle}{\emph{Proceedings of the 29th annual
  conference on Computer graphics and interactive techniques}}.
  \bibinfo{pages}{501--508}.
\newblock


\bibitem[\protect\citeauthoryear{Rose, Cohen, and Bodenheimer}{Rose
  et~al\mbox{.}}{1998}]%
        {rose1998verbs}
\bibfield{author}{\bibinfo{person}{Charles Rose}, \bibinfo{person}{Michael~F
  Cohen}, {and} \bibinfo{person}{Bobby Bodenheimer}.}
  \bibinfo{year}{1998}\natexlab{}.
\newblock \showarticletitle{Verbs and adverbs: Multidimensional motion
  interpolation}.
\newblock \bibinfo{journal}{\emph{IEEE Computer Graphics and Applications}}
  \bibinfo{volume}{18}, \bibinfo{number}{5} (\bibinfo{year}{1998}),
  \bibinfo{pages}{32--40}.
\newblock


\bibitem[\protect\citeauthoryear{Rose, Guenter, Bodenheimer, and Cohen}{Rose
  et~al\mbox{.}}{1996}]%
        {rose1996efficient}
\bibfield{author}{\bibinfo{person}{Charles Rose}, \bibinfo{person}{Brian
  Guenter}, \bibinfo{person}{Bobby Bodenheimer}, {and}
  \bibinfo{person}{Michael~F Cohen}.} \bibinfo{year}{1996}\natexlab{}.
\newblock \showarticletitle{Efficient generation of motion transitions using
  spacetime constraints}. In \bibinfo{booktitle}{\emph{Proceedings of the 23rd
  annual conference on Computer graphics and interactive techniques}}.
  \bibinfo{pages}{147--154}.
\newblock


\bibitem[\protect\citeauthoryear{Safonova and Hodgins}{Safonova and
  Hodgins}{2007}]%
        {safonova2007construction}
\bibfield{author}{\bibinfo{person}{Alla Safonova} {and}
  \bibinfo{person}{Jessica~K Hodgins}.} \bibinfo{year}{2007}\natexlab{}.
\newblock \showarticletitle{Construction and optimal search of interpolated
  motion graphs}.
\newblock In \bibinfo{booktitle}{\emph{ACM SIGGRAPH 2007 papers}}.
  \bibinfo{pages}{106--es}.
\newblock


\bibitem[\protect\citeauthoryear{Shaham, Dekel, and Michaeli}{Shaham
  et~al\mbox{.}}{2019}]%
        {shaham2019singan}
\bibfield{author}{\bibinfo{person}{Tamar~Rott Shaham}, \bibinfo{person}{Tali
  Dekel}, {and} \bibinfo{person}{Tomer Michaeli}.}
  \bibinfo{year}{2019}\natexlab{}.
\newblock \showarticletitle{SinGAN: Learning a generative model from a single
  natural image}. In \bibinfo{booktitle}{\emph{Proceedings of the IEEE/CVF
  International Conference on Computer Vision}}. \bibinfo{pages}{4570--4580}.
\newblock


\bibitem[\protect\citeauthoryear{Shocher, Bagon, Isola, and Irani}{Shocher
  et~al\mbox{.}}{2019}]%
        {shocher2019ingan}
\bibfield{author}{\bibinfo{person}{Assaf Shocher}, \bibinfo{person}{Shai
  Bagon}, \bibinfo{person}{Phillip Isola}, {and} \bibinfo{person}{Michal
  Irani}.} \bibinfo{year}{2019}\natexlab{}.
\newblock \showarticletitle{Ingan: Capturing and retargeting the" dna" of a
  natural image}. In \bibinfo{booktitle}{\emph{Proceedings of the IEEE/CVF
  International Conference on Computer Vision}}. \bibinfo{pages}{4492--4501}.
\newblock


\bibitem[\protect\citeauthoryear{Starke, Zhao, Komura, and Zaman}{Starke
  et~al\mbox{.}}{2020}]%
        {starke2020local}
\bibfield{author}{\bibinfo{person}{Sebastian Starke}, \bibinfo{person}{Yiwei
  Zhao}, \bibinfo{person}{Taku Komura}, {and} \bibinfo{person}{Kazi Zaman}.}
  \bibinfo{year}{2020}\natexlab{}.
\newblock \showarticletitle{Local motion phases for learning multi-contact
  character movements}.
\newblock \bibinfo{journal}{\emph{ACM Transactions on Graphics (TOG)}}
  \bibinfo{volume}{39}, \bibinfo{number}{4} (\bibinfo{year}{2020}),
  \bibinfo{pages}{54--1}.
\newblock


\bibitem[\protect\citeauthoryear{Starke, Zhao, Zinno, and Komura}{Starke
  et~al\mbox{.}}{2021}]%
        {starke2021neural}
\bibfield{author}{\bibinfo{person}{Sebastian Starke}, \bibinfo{person}{Yiwei
  Zhao}, \bibinfo{person}{Fabio Zinno}, {and} \bibinfo{person}{Taku Komura}.}
  \bibinfo{year}{2021}\natexlab{}.
\newblock \showarticletitle{Neural animation layering for synthesizing martial
  arts movements}.
\newblock \bibinfo{journal}{\emph{ACM Transactions on Graphics (TOG)}}
  \bibinfo{volume}{40}, \bibinfo{number}{4} (\bibinfo{year}{2021}),
  \bibinfo{pages}{1--16}.
\newblock


\bibitem[\protect\citeauthoryear{Tanco and Hilton}{Tanco and Hilton}{2000}]%
        {tanco2000realistic}
\bibfield{author}{\bibinfo{person}{Luis~Molina Tanco} {and}
  \bibinfo{person}{Adrian Hilton}.} \bibinfo{year}{2000}\natexlab{}.
\newblock \showarticletitle{Realistic synthesis of novel human movements from a
  database of motion capture examples}. In
  \bibinfo{booktitle}{\emph{Proceedings Workshop on Human Motion}}. IEEE,
  \bibinfo{pages}{137--142}.
\newblock


\bibitem[\protect\citeauthoryear{Taylor and Hinton}{Taylor and Hinton}{2009}]%
        {taylor2009factored}
\bibfield{author}{\bibinfo{person}{Graham~W Taylor} {and}
  \bibinfo{person}{Geoffrey~E Hinton}.} \bibinfo{year}{2009}\natexlab{}.
\newblock \showarticletitle{Factored conditional restricted Boltzmann machines
  for modeling motion style}. In \bibinfo{booktitle}{\emph{Proceedings of the
  26th annual international conference on machine learning}}.
  \bibinfo{pages}{1025--1032}.
\newblock


\bibitem[\protect\citeauthoryear{{Truebones Motions Animation
  Studios}}{{Truebones Motions Animation Studios}}{2022}]%
        {truebones}
\bibfield{author}{\bibinfo{person}{{Truebones Motions Animation Studios}}.}
  \bibinfo{year}{2022}\natexlab{}.
\newblock \bibinfo{title}{Truebones}.
\newblock
\newblock
\urldef\tempurl%
\url{https://truebones.gumroad.com/}
\showURL{%
\tempurl}
\newblock
\shownote{Accessed: 2022-1-15.}


\bibitem[\protect\citeauthoryear{Villegas, Yang, Ceylan, and Lee}{Villegas
  et~al\mbox{.}}{2018}]%
        {villegas2018neural}
\bibfield{author}{\bibinfo{person}{Ruben Villegas}, \bibinfo{person}{Jimei
  Yang}, \bibinfo{person}{Duygu Ceylan}, {and} \bibinfo{person}{Honglak Lee}.}
  \bibinfo{year}{2018}\natexlab{}.
\newblock \showarticletitle{Neural kinematic networks for unsupervised motion
  retargetting}. In \bibinfo{booktitle}{\emph{Proceedings of the IEEE
  Conference on Computer Vision and Pattern Recognition}}.
  \bibinfo{pages}{8639--8648}.
\newblock


\bibitem[\protect\citeauthoryear{Wang, Fleet, and Hertzmann}{Wang
  et~al\mbox{.}}{2007}]%
        {wang2007gaussian}
\bibfield{author}{\bibinfo{person}{Jack~M Wang}, \bibinfo{person}{David~J
  Fleet}, {and} \bibinfo{person}{Aaron Hertzmann}.}
  \bibinfo{year}{2007}\natexlab{}.
\newblock \showarticletitle{Gaussian process dynamical models for human
  motion}.
\newblock \bibinfo{journal}{\emph{IEEE transactions on pattern analysis and
  machine intelligence}} \bibinfo{volume}{30}, \bibinfo{number}{2}
  (\bibinfo{year}{2007}), \bibinfo{pages}{283--298}.
\newblock


\bibitem[\protect\citeauthoryear{Wei, Min, and Chai}{Wei et~al\mbox{.}}{2011}]%
        {wei2011physically}
\bibfield{author}{\bibinfo{person}{Xiaolin Wei}, \bibinfo{person}{Jianyuan
  Min}, {and} \bibinfo{person}{Jinxiang Chai}.}
  \bibinfo{year}{2011}\natexlab{}.
\newblock \showarticletitle{Physically valid statistical models for human
  motion generation}.
\newblock \bibinfo{journal}{\emph{ACM Transactions on Graphics (TOG)}}
  \bibinfo{volume}{30}, \bibinfo{number}{3} (\bibinfo{year}{2011}),
  \bibinfo{pages}{1--10}.
\newblock


\bibitem[\protect\citeauthoryear{Wiley and Hahn}{Wiley and Hahn}{1997}]%
        {wiley1997interpolation}
\bibfield{author}{\bibinfo{person}{Douglas~J Wiley} {and}
  \bibinfo{person}{James~K Hahn}.} \bibinfo{year}{1997}\natexlab{}.
\newblock \showarticletitle{Interpolation synthesis of articulated figure
  motion}.
\newblock \bibinfo{journal}{\emph{IEEE Computer Graphics and Applications}}
  \bibinfo{volume}{17}, \bibinfo{number}{6} (\bibinfo{year}{1997}),
  \bibinfo{pages}{39--45}.
\newblock


\bibitem[\protect\citeauthoryear{Ye and Liu}{Ye and Liu}{2010}]%
        {ye2010synthesis}
\bibfield{author}{\bibinfo{person}{Yuting Ye} {and} \bibinfo{person}{C~Karen
  Liu}.} \bibinfo{year}{2010}\natexlab{}.
\newblock \showarticletitle{Synthesis of responsive motion using a dynamic
  model}. In \bibinfo{booktitle}{\emph{Computer Graphics Forum}},
  Vol.~\bibinfo{volume}{29}. Wiley Online Library, \bibinfo{pages}{555--562}.
\newblock


\bibitem[\protect\citeauthoryear{Zhang, Starke, Komura, and Saito}{Zhang
  et~al\mbox{.}}{2018}]%
        {zhang2018mode}
\bibfield{author}{\bibinfo{person}{He Zhang}, \bibinfo{person}{Sebastian
  Starke}, \bibinfo{person}{Taku Komura}, {and} \bibinfo{person}{Jun Saito}.}
  \bibinfo{year}{2018}\natexlab{}.
\newblock \showarticletitle{Mode-adaptive neural networks for quadruped motion
  control}.
\newblock \bibinfo{journal}{\emph{ACM Transactions on Graphics (TOG)}}
  \bibinfo{volume}{37}, \bibinfo{number}{4} (\bibinfo{year}{2018}),
  \bibinfo{pages}{1--11}.
\newblock


\bibitem[\protect\citeauthoryear{Zhao, Normoyle, Khanna, and Safonova}{Zhao
  et~al\mbox{.}}{2009}]%
        {zhao2009automatic}
\bibfield{author}{\bibinfo{person}{Liming Zhao}, \bibinfo{person}{Aline
  Normoyle}, \bibinfo{person}{Sanjeev Khanna}, {and} \bibinfo{person}{Alla
  Safonova}.} \bibinfo{year}{2009}\natexlab{}.
\newblock \showarticletitle{Automatic construction of a minimum size motion
  graph}. In \bibinfo{booktitle}{\emph{Proceedings of the 2009 ACM
  SIGGRAPH/Eurographics symposium on Computer animation}}.
  \bibinfo{pages}{27--35}.
\newblock


\bibitem[\protect\citeauthoryear{Zhou, Barnes, Lu, Yang, and Li}{Zhou
  et~al\mbox{.}}{2019}]%
        {zhou2019continuity}
\bibfield{author}{\bibinfo{person}{Yi Zhou}, \bibinfo{person}{Connelly Barnes},
  \bibinfo{person}{Jingwan Lu}, \bibinfo{person}{Jimei Yang}, {and}
  \bibinfo{person}{Hao Li}.} \bibinfo{year}{2019}\natexlab{}.
\newblock \showarticletitle{On the continuity of rotation representations in
  neural networks}. In \bibinfo{booktitle}{\emph{Proceedings of the IEEE/CVF
  Conference on Computer Vision and Pattern Recognition}}.
  \bibinfo{pages}{5745--5753}.
\newblock


\bibitem[\protect\citeauthoryear{Zhou, Li, Xiao, He, Huang, and Li}{Zhou
  et~al\mbox{.}}{2018}]%
        {zhou2018auto}
\bibfield{author}{\bibinfo{person}{Yi Zhou}, \bibinfo{person}{Zimo Li},
  \bibinfo{person}{Shuangjiu Xiao}, \bibinfo{person}{Chong He},
  \bibinfo{person}{Zeng Huang}, {and} \bibinfo{person}{Hao Li}.}
  \bibinfo{year}{2018}\natexlab{}.
\newblock \showarticletitle{Auto-Conditioned Recurrent Networks for Extended
  Complex Human Motion Synthesis}. In \bibinfo{booktitle}{\emph{International
  Conference on Learning Representations}}.
\newblock


\end{thebibliography}
